\documentclass[preprint,3p]{elsarticle}

\usepackage{amsmath}
\usepackage{amssymb}
\usepackage{graphicx}
\usepackage{epsfig}
\usepackage{color}
\usepackage{url}
\usepackage{times}
\usepackage{bm}

\newcommand{\beq}{\begin{equation}}
\newcommand{\eeq}{\end{equation}}
\newcommand{\bea}{\begin{eqnarray}}
\newcommand{\eea}{\end{eqnarray}}
\newcommand{\bit}{\begin{itemize}}
\newcommand{\eit}{\end{itemize}}
\newcommand{\ben}{\begin{enumerate}}
\newcommand{\een}{\end{enumerate}}
\newcommand{\nn}{\nonumber}

\newcommand{\coord}{\tau, \sigma, \theta}
\newcommand{\gencoord}{\tau, x^a}

\newcommand{\spacecoord}{\sigma, \theta}

\font\tenscr=rsfs10 scaled1100
\font\sevenscr=rsfs7 % scaled \magstep1
\font\fivescr=rsfs5 % scaled \magstep1
\skewchar\tenscr='177
\skewchar\sevenscr='177
\skewchar\fivescr='177
\newfam\scrfam
\textfont\scrfam=\tenscr
\scriptfont\scrfam=\sevenscr
\scriptscriptfont\scrfam=\fivescr

\def\scri{{{\fam\scrfam I}^+}}

\newcommand{\TPI}{\address{Theoretisch-Physikalisches Institut,
    Friedrich-Schiller-Universit\"at Jena,\\ Max-Wien-Platz 1,
          D-07743 Jena, Germany}}

\begin{document}
\title{Axisymmetric  fully spectral code for hyperbolic equations}
\author{Rodrigo Panosso Macedo} 
\ead{rodrigo.panosso-macedo@uni-jena.de}

\author{Marcus Ansorg} 
\ead{marcus.ansorg@uni-jena.de}
\TPI
\date{\today}

\begin{abstract}
We present a fully pseudo-spectral scheme to solve axisymmetric hyperbolic equations of second order. With the Chebyshev polynomials as basis functions,  the numerical grid is based on the Lobbato (for two spatial directions) and Radau (for the time direction) collocation points. The method solves two issues of previous algorithms which were restricted to one spatial dimension, namely,  (i) the inversion of a dense matrix and (ii) the acquisition  of a sufficiently good initial-guess for non-linear systems of equations. For the first issue, we use the iterative bi-conjugate gradient stabilized method, which we equip with a pre-conditioner based on a singly diagonally implicit Runge-Kutta ("SDIRK"-) method. In this paper, the SDIRK-method is also used to solve issue (ii). The numerical solutions are correct up to machine precision and we do not observe any restriction concerning the time step in comparison with the spatial resolution. As an application, we solve general-relativistic wave equations on a black-hole space-time in so-called hyperboloidal slices and reproduce some recent results available in the literature.
\end{abstract}

\maketitle

%%%%%%%%%%%%%%%%%%%%%%%%%%%%%%%%%%%%%%%%%%%%%%%%
\section{Introduction}
There is no need to emphasize the central role that partial differential equations (PDEs) play in physics, since probably most processes in nature can be modeled by them. While the mathematical theory of linear PDEs is a well established field and exact solutions are available in many cases, treating a system of nonlinear PDEs analytically is a much more complicated task, where rarely does one get an exact solution. In such cases, numerical simulation is a very useful tool to approach the problem. However, the numerical solution to the PDEs system is not exact. Rather, it is beset with errors introduced by the chosen numerical method.   

Even though there is a wide range of numerical methods to treat PDEs, our interests lie in those which render highly accurate solutions, ideally close to machine precision. In this context, (pseudo-)spectral methods are probably the best choice, as they have the remarkable capability of providing exponential convergence rate when the underlying problem admits a regular solution~\cite{Boyd00,canuto_2006_smf}. 

Spectral methods have been widely used for systems of {\em elliptic} PDEs~\cite{Boyd00,canuto_2006_smf}. For time-dependent problems though, usually the spatial and time directions are treated differently (see e.g.~\cite{hesthaven2007} for a discussion on the use of spectral methods for time-dependent problems). Typically, spectral methods are restricted to the spatial grid, while the time evolution is performed with a standard time integrator (for instance, the well known explicit 4th order Runge-Kutta scheme). Apart from the loss of the solution's accuracy, in many cases such combined methods have to satisfy the so-called Courant-Friedrichs-Lewy condition (CFL condition) which imposes a restriction on the time step according to the spatial grid size.  A way to overcome these caveats is to apply a spectral method to {\em all} directions, i.e.~to both space {\em and} time. To the best of our knowledge, a first study along this line was performed for {\em parabolic} PDEs \cite{Ierley199288}, while a fully spectral code for hyperbolic equations was first proposed in~\cite{Hennig:2008af}. These works were restricted to problems with one spatial and one time dimension. Recently, a first step towards a fully spectral code in higher dimensions was presented in~\cite{Petri2014} for advection equations.

In this work, we extend the results of \cite{Hennig:2008af,Ansorg:2011xg, Hennig:2012zx} and introduce a fully spectral code for axisymmetric hyperbolic PDEs. In particular, we address two limitations of the code presented in \cite{Hennig:2012zx}. The first issue concerns the inversion of a large dense matrix, which is known to be a delicate aspect inside a fully spectral solver (see e.g.~\cite{Grandclement:2009ju}). In higher dimensions (for our axisymmetric code: two spatial dimensions and one time dimension, henceforth "2+1"), the computational costs of a direct method (such as LU-decomposition) become prohibitive (as reported, for instance, in \cite{Petri2014}). The second issue arises since the treatment of nonlinear equations by means of the Newton-Raphson scheme (which is frequently used in the realm of spectral methods) requires an adequate initial-guess for the solution. The examples shown in \cite{Hennig:2012zx} are small perturbations of an explicitly known solution. For stronger perturbations, the method again becomes expensive, as one would have to go through many intermediate steps with gradually increasing perturbation parameter until the desired situation is reached. Even more seriously, in cases in which no corresponding approximate solution can be identified, the method might not work at all, as no initial-guess would be available.

In order to solve the first issue, we invert the dense matrices appearing within the spectral code in terms of the iterative "bi-conjugate gradient stabilized method" (BiCGStab)~\cite{Barrett93}. In particular, we endow the BiCGStab method with a pre-conditioner based on a "singly diagonally implicit Runge-Kutta method" (SDIRK), which reduces significantly the number of iterations needed for convergence inside the BiCGStab-method. The SDIRK-method is also the algorithm of our choice to solve the second issue \footnote{In this case, the procedure can be seen as follows: given initial data and boundary conditions, we first solve the PDE system with an implicit Runge-Kutta integrator in time and spectral methods for spatial coordinates; this first solution is then refined by the use of spectral methods in time. We mention however that (i) for linear equations such a preliminary step is not necessary and (ii) for non-linear equations, the initial guess could alternatively be determined by other means.}.

Note that a delicate issue arises when the spatial domain is unbounded (for instance, with a radius coordinate $r\in [0, \infty)$). In principle, one could introduce artificial boundaries at large, but finite radii on the numerical grid. However, in order to complete the mathematical problem, one needs extra conditions which are compatible with the differential equations and with the underlying physical scenario. Unfortunately, it is a very difficult task to fully control the influence from such artificial boundaries on the numerical solution. Especially, pseudo-spectral methods are particularly sensitive to this feature, and the simulation is bound to break down after a few time steps, when errors originating from the boundary accumulate. 

For such unbounded domains, a common strategy is the compactification of the spatial domain via the introduction of a suitable coordinate system. This approach is not just a clever trick to treat the equations numerically, but it is an active line of research within General Relativity. In \cite{Penrose2}, Penrose introduced the concept of conformal infinity, which brings so-called future null infinity $\scri$ (which is the set of points which are approached asymptotically by light rays and gravitational waves) to a finite coordinate value. In this manner, the conformal concept permits the inclusion of this surface in the numerical grid and it removes the necessity of imposing artificially boundary conditions. Generally speaking, the physical space-time metric ${g}_{\mu \nu}$ is rescaled by a conformal factor $\Omega$ and one works with the regular {\em conformal} metric $\tilde{g}_{\mu \nu}=\Omega^{2} {g}_{\mu \nu}$ (see \cite{Frauendiener04} for a review). In the compactified "unphysical" space-time endowed with the conformal metric, future null infinity is described simply by the hypersurface $\Omega_{|\scri} = 0$. In this context, space-like surfaces extending up to $\scri$ are referred to as hyperboloidal slices. 

From the physical point of view, the inclusion of $\scri$ in the numerical grid allows one to precisely extract the gravitational wave content of the space-time. However, the conformal approach has the drawback of introducing singular terms ($\sim \Omega^{-n}$ for some integer $n$) into the field equations. For relativistic gravitational fields, Friedrich~\cite{Friedrich:1983} reformulated the Einstein equations in terms of the conformal metric in such a way that they are manifestly regular at $\scri$. Regarding the practical purpose, the system of equations turned out to be rather complicated for the numerical treatment (but see \cite{frauendiener2002conformal}, for a complete discussion on the theoretical and numerical development of this formalism).  Recently, Moncrief and Rinne~\cite{Moncrief:2008ie} showed that the apparently singular boundary terms appearing in the hyperboloidal concept can be explicitly evaluated at $\scri$. Their scheme is based on a constrained and gauge fixing formulation of the field equation and, as a result, stable dynamical numerical evolutions were presented in~\cite{Rinne:2009qx,Rinne:2013qc}. An alternative approach was suggested in~\cite{Zenginoglu:2008pw} where a free unconstrained evolution of the field equations in terms of the generalized harmonic gauge~\cite{Friedrich:2000qv} is described. 

In a future research effort we intend to numerically solve the Einstein equations in such a harmonic formalism, in which they form a system of wave equations. The work presented in this article is a pre-intermediate step toward this goal and we show, in particular, that our fully spectral method can handle equations with singular terms and that it is suitable to deal with the conformal approach. Another feature of our code is the capability of solving problems with {\em free} boundaries such as the dynamics of a spherically symmetric, oscillating, self-gravitating star whose spatial location of its surface is unknown at the outset but needs to be determined simultaneously with the evolution of the gravitational field and the star's density and velocity fields.

The paper is organized as follows. Section \ref{sec:NumMeth} introduces the numerical scheme. It describes both the fully spectral and the SDIRK methods and it ends with a simple example that illustrates the application of the SDIRK method in a free-boundary problem. In section \ref{sec:Applications} we apply the scheme to two different problems. At first we reproduce the results of \cite{Hennig:2012zx} regarding the computation of the dynamics of a spherically symmetric, oscillating, self-gravitating star. We focus, in  particular, on the improvement of performance. In the second application we expand the scheme to (2+1)-dimensions and solve the axisymmetric wave equation on a space-time background of a rotating black hole. Finally, we present our conclusions.

We finish the introduction with a comment on the notation: uppercase Latin letters ($A,B,...$) run over field variables;  lower case Latin letter (in particular $i,j,k$) are reserved for the numerical grid points; lower case Latin letter in parenthesis $(i), (j), ...$ specify quantities of a Runge-Kutta method; letters from the Greek alphabet ($\mu, \nu...$) designate the components of a tensor in a given coordinate system.

%%%%%%%%%%%%%%%%%%%%%%%%%%%%%%%%%%%%%%%%%%%%%%%%%%%%%%%%%
\section{Numerical Methods}\label{sec:NumMeth}
In this section we describe the fully spectral method to solve axisymmetric hyperbolic PDEs of second order. In its most general concept, the scheme is adapted to handle nonlinear equations, with numerical boundaries that are possibly not known a priori. In this way, we assume a system of $n_{\rm var}$ variables $X^A$ described by equations written implicitly in the generic compact form
\beq
\label{eq:SpecEqGenForm}
E^A\left(\tau, \left\{ X^B\right\}, \left\{ \frac{\partial X^B}{\partial\tau} \right\}, \left\{ \frac{\partial^2 X^B}{\partial\tau^2}\right\} \right) = 0.
\eeq
Here, $\tau$ is the time coordinate whereas $\sigma$ and $\theta$ refer to radial and polar-angle coordinates of a spherical-like coordinate system (explicit dependency on $\sigma$ and $\theta$ was suppressed in (\ref{eq:SpecEqGenForm})). The variables $X^A$ represent not only the $n_{\rm fields}$ axisymmetric fields defined for $(\tau, \sigma, \theta) \in [\tau_{\rm a}, \tau_{\rm b}]\times[\sigma_{\rm a}, \sigma_{\rm b}]\times[0,\pi]$, whose dynamics are dictated by hyperbolic PDEs. The notation also includes $n_{\rm boundary}$ functions defined for $(\tau, \theta) \in [\tau_{\rm a}, \tau_{\rm b}]\times[0,\pi]$ describing values at the boundaries in a free-boundary problem. In this way, the total number of variables amounts to $n_{\rm var} = n_{\rm fields} +n_{\rm boundary}$. 

To simplify the notation, we shall write the variable dependence as $X^A = X^A(\tau,x^a)$. The spatial variables $x^a$ stand for $\{\sigma, \theta\}$ if $X^A$ describes the axisymmetric fields, while $x^a$ corresponds to $\theta$ if $X^A$ describes the boundaries or boundary values in the free-boundary problem. When needed, the dependence will be explicitly stated. Similarly, the functions $E^A$ incorporate both the hyperbolic PDEs as well as boundary conditions that need to be added in a free-boundary problem to complete the formulation (in the explicit example presented below, this point will become clear). Note that the notation $\left\{ \bullet \right\}$ comprises all {\em spatial} derivatives of the argument. That is to say that the arguments in (\ref{eq:SpecEqGenForm}) stand for
\bea
	\left\{X^B\right\}&=&\left\{X^1, X^1_{,a}, X^1_{,ab},X^2, X^2_{,a}, X^2_{,ab},\ldots, X^{n_{\rm nvar}}, X^{n_{\rm nvar}}_{,\theta}, X^{n_{\rm nvar}}_{,\theta\theta}\right\}\\
\left\{ \frac{\partial X^B}{\partial\tau} \right\}&=& \left\{
	\left(\frac{\partial X^1}{\partial\tau}\right), 
	\left(\frac{\partial X^1}{\partial\tau}\right)_{,a},\ldots, 
	\left(\frac{\partial X^{n_{\rm var}}}{\partial\tau}\right), 
	\left(\frac{\partial X^{n_{\rm var}}}{\partial\tau}\right)_{,\theta}\right\}\\
\left\{ \frac{\partial^2 X^B}{\partial\tau^2} \right\}&=& \left\{
	\left(\frac{\partial^2 X^1}{\partial\tau^2}\right),\ldots, 
	\left(\frac{\partial^2 X^{n_{\rm var}}}{\partial\tau^2}\right)\right\}
\eea
The notation $\bullet_{,a}$ denotes partial derivatives with respect to the spatial variables $(x^a)=(\sigma, \theta)$. Note that for second order PDEs no terms $[(\partial X^B)/(\partial\tau)]_{,ab}$ and $[(\partial^2 X^B)/(\partial\tau^2)]_{,a}$ appear. Besides the possible boundary conditions we need to impose {\em initial data} to render a unique solution and to complete the formulation of the problem. For the second-order Cauchy-type initial value problem (\ref{eq:SpecEqGenForm}) we require initial data.
\beq
\label{eq:InitialData}
X^A(\tau_{\rm a},x^a) = X_{\rm initial}^A(x^a),\quad 
\frac{\partial X^A}{\partial\tau} (\tau_{\rm a},x^a) = Y_{\rm initial}^A(x^a),\quad A=1,\ldots,n_{\rm var},
\eeq
where the given functions $X_{\rm initial}^A$ and $Y_{\rm initial}^A$ describe the fields and boundary functions as well as their first time derivatives at the initial point in time, $\tau=\tau_{\rm a}$.

\begin{figure*}[h!]
\begin{center}
\includegraphics[width=8.1cm]{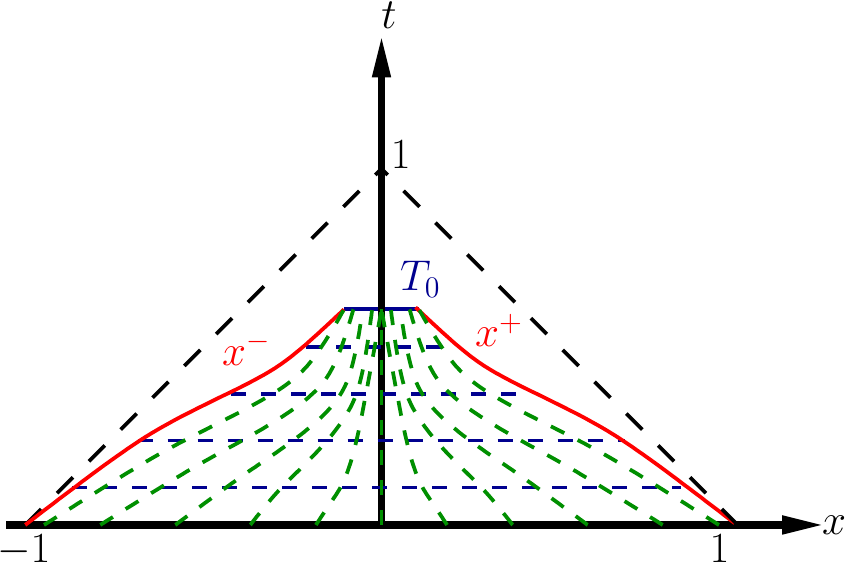}
\includegraphics[width=8.1cm]{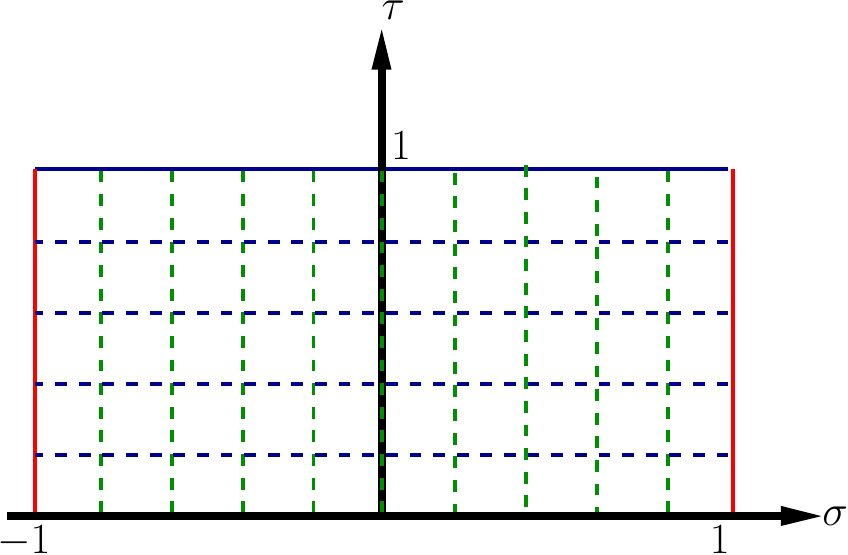}
\end{center}
\caption{Coordinate transformation given by eqs. (\ref{eq:WaveEqCoordTrans}) for the free boundary problem formulation of the (1+1)-wave equation. The free boundaries $x^\pm$ are obtained for the constant coordinate values $\sigma=\pm 1$.
 }
\label{fig:WaveEqDom}
\end{figure*}

We illustrate this abstract formulation with an explicit example. Consider, as a specific free boundary hyperbolic problem, the wave equation 
\beq
\label{eq:waveEq}
 -\phi_{,tt} + \phi_{,xx} = 0
\eeq
in time (coordinate $t$) and one spatial dimension (coordinate $x$). Let us assume that we seek the solution of this equation in a domain which is parametrically given by
\beq
\label{eq:WaveEqCoordTrans}
 \left\{(x,t)\in \mathbb{R}^2\,\Big|\,
		t = T_0\tau,\, 
		x = \frac{1}{2}\left[ (1-\sigma) x^-(\tau) + (1+\sigma)x^+(\tau)\right]\,,\,\,
		(\sigma,\tau)\in[-1,1]\times[0,1]\right\},
\eeq 
for some prescribed $T_0<1$, see fig.\ref{fig:WaveEqDom}. Here the domain boundaries $x^\pm$ are supposed to be functions in $\tau$ which are {\em unknown} from the outset. Rather, in addition to the requirement of the validity of the wave equation we impose Dirichlet-type boundary conditions, i.e.
\beq\label{eq:WaveEqBoundCond}\phi(\tau,\sigma=\pm 1)=\phi^\pm(\tau) \eeq
for some prescribed boundary values $\phi^\pm=\phi^\pm(\tau)$. Note that, for the solubility of this problem, it is necessary that the solution obeys ${dx^-}/{d\tau}\geq T_0$ and ${dx^+}/{d\tau}\leq-T_0$, since these conditions guarantee that the characteristics of the wave equation always point outward the numerical domain and hence no further boundary conditions at $\sigma=\pm 1$ are required.

\noindent
We express the wave equation (\ref{eq:waveEq}) in the coordinates $(\tau,\sigma)$,
\bea
&& \left(x^+-x^-\right)^2\ddot{\phi}- \left\{4T_0^2- \left[(1-\sigma)\dot{x}^- + (1+\sigma)\dot{x}^+\right]^2 \right\}\phi'' \nonumber \\
 &&+ 2\left[(1-\sigma)\dot{x}^- + (1+\sigma)\dot{x}^+ \right] \left[ \phi'\left( \dot{x}^+ - \dot{x}^- \right) - \dot{\phi}'(x^+-x^-)\right]    \nonumber \\
&& - (x^+-x^-)\left[(1-\sigma)\ddot{x}^- + (1+\sigma)\ddot{x}^+\right]\phi'= 0, \label{eq:waveEq2}
\eea 
where we have written $\phi=\phi(\tau,\sigma)$ and $\dot{\phi}=\partial \phi/\partial \tau$ (accordingly $\dot{x}^\pm =  d x^\pm/d\tau$ ) and $\phi' = \partial \phi/\partial \sigma$. From a numerical point of view (see discussion below), it will turn out to be advantageous to rewrite the boundary conditions (\ref{eq:WaveEqBoundCond}) in the form
\beq\label{eq:ddot_WaveEqBoundCond}\ddot\phi(\tau,\sigma=\pm 1)=\ddot\phi^\pm(\tau), \eeq
which, by virtue of the requirement of a complete set of initial data, 
\beq \label{eq:WaveEqBoundCond_ID}
\{\phi(0,\sigma),\dot\phi(0,\sigma), x^\pm(0), \dot x^\pm(0)\},
\eeq
is equivalent to (\ref{eq:WaveEqBoundCond}). In this manner we obtain the  hyperbolic equation (\ref{eq:waveEq2}) with unknown functions $\{\phi(\tau,\sigma), x^\pm(\tau)\}$ which has to be solved for $(\sigma,\tau)\in[0,1]\times[-1,1]$ subject to the boundary conditions (\ref{eq:ddot_WaveEqBoundCond}). 

\noindent
In the simple example considered, an analytic solution is available in the form 
\bea
\phi_{\rm exact}(\tau, \sigma) &=& f(v(\tau,\sigma)) + g(u(\tau,\sigma)), \quad {\rm with} \nn \\
u(\tau,\sigma)&=&t(\tau)-x(\tau,\sigma) \quad {\rm and} \label{eq:phi_exact} \\
v(\tau,\sigma)&=&t(\tau)+x(\tau,\sigma), \nonumber
\eea
where we read $t(\tau)$ and $x(\tau,\sigma)$ from (\ref{eq:WaveEqCoordTrans}).
So, for the validation of the method (and the corresponding code) we can prepare a setup for the free boundary hyperbolic problem by (i) prescribing $T_0<1$ and (ii) taking some arbitrary functions $f$ and $g$ as well as boundary curves $x_{\rm exact}^\pm(t/T_0)$. From these functions we compute 
\[\ddot\phi^\pm(\tau)=\ddot\phi_{\rm exact}(\tau, \pm 1)\]
as well as the initial data for $\phi$,
\[\phi(0,\sigma) = \phi_{\rm exact}(0, \sigma),\qquad \dot\phi(0,\sigma) = \dot\phi_{\rm exact}(0, \sigma),\]
and those for $x^\pm$,
\[x^\pm(0)=x_{\rm exact}^\pm(0),\qquad\dot x^\pm(0)=\dot x_{\rm exact}^\pm(0).\]
At this point, we may feed our algorithm to solve (\ref{eq:waveEq2}) together with (\ref{eq:ddot_WaveEqBoundCond}) with the data 
\beq
\label{eq:phi_exact_ID}
\{\ddot\phi^\pm(\tau),\underbrace{\phi_{\rm exact}(0,\sigma),\dot\phi_{\rm exact}(0,\sigma), x_{\rm exact}^\pm(0), \dot x_{\rm exact}^\pm(0)}_{\mbox{initial data}}\}.
\eeq 
The hope is that eventually the solver provides us with an accurate approximation of the original functions $\phi_{\rm exact}(\tau,\sigma)$ 
{\em and} $x_{\rm exact}^\pm(\tau)$ for all $(\sigma,\tau)\in[0,1]\times[-1,1]$.

\noindent
We now may cast eqs. (\ref{eq:waveEq2}) and (\ref{eq:ddot_WaveEqBoundCond}) into the form of (\ref{eq:SpecEqGenForm}). To this end we identify the variables $\phi,$ $x^+$ and $x^-$ as follows:
\[
	\begin{array}{ccccccccc}
	  X^1 &=& \phi, & Y^1&=&\dot{\phi}, &  Z^1 &=& \ddot{\phi}, \\[3mm]
	  X^2 &=& x^+, & Y^2&=&\dot{x}^+, &  Z^2 &=& \ddot{x}^+,\\[3mm]
	  X^3 &=& x^-, & Y^3&=&\dot{x}^-, &  Z^3 &=& \ddot{x}^-, 
	\end{array}
\]
and write
\bea
E^1\left(\tau, \{X^A\}, \{Y^A\}, \{Z^A\}\right) &=& \left(X^2-X^3\right)^2Z^1 - 
\left\{4T_0^2- \left[(1-\sigma)Y^3 + (1+\sigma)Y^2\right]^2\right\}\left(X^1\right)''\nonumber \\
&& + 2\left[(1-\sigma)Y^3 + (1+\sigma)Y^2 \right] \left[ \left(X^1\right)'\left( Y^2 - Y^3 \right) - \left(Y^1\right)'(X^2-X^3)\right]\nonumber \\
&&-(X^2-X^3)\left[(1-\sigma)Z^3 + (1+\sigma)Z^2\right](X^1)'  \label{eq:WaveEqFreeBound_F1} \\
E^2\left(\tau, \{X^A\}, \{Y^A\}, \{Z^A\}\right) &=& \left. Z^1\right|_{\sigma = -1} - \ddot \phi^-(\tau),\label{eq:WaveEqFreeBound_F2} \\
E^3\left(\tau, \{X^A\}, \{Y^A\}, \{Z^A\}\right) &=& \left. Z^1\right|_{\sigma = +1} - \ddot \phi^+(\tau). \label{eq:WaveEqFreeBound_F3}
\eea

\vspace*{5mm}
We would like to put emphasis on the fact that, in their most generic form, eqs.~(\ref{eq:SpecEqGenForm}) might contain singular terms (for instance, going as $\sim(\sin{\theta})^{-n}$ or as $\sim\sigma^{-n}$ when $\sigma_{\rm a}=0$, for some integer $n$). Usually, such terms appear with the choice of particular coordinate systems and they naturally arise within the formalism of conformal infinity in General Relativity. One can, in principle, divide the equation by the corresponding singular term (say $\sim\sigma^{-n}$). However, at the surface $\sigma=0$, the term proportional to ${\partial^2X^A}/{\partial\tau^2}$ vanishes then and an explicit time integrator (such as the famous 4th-order Runge-Kutta method) would not be suitable for numerically evolving eqs.~(\ref{eq:SpecEqGenForm}). Also, the presence of singularities implies regularity conditions for the solution and, in particular, the initial data must be chosen accordingly. 

In the general situation, we proceed with the treatment of the eqs.~(\ref{eq:SpecEqGenForm}) as described in ~\cite{Hennig:2008af}. That is, 
for given regular initial data $X^A_{\rm initial}(x^a) = X^A(\tau_{\rm a}, x^a)$ and $Y^A_{\rm initial}(x^a) = \partial_\tau X^A(\tau_{\rm a}, x^a)$, we introduce auxiliary functions $U^A(\tau, x^a)$ such that the original quantities are expressed as follows:
\bea
\label{eq:newField}
X^A(\gencoord) &=& X^A_{\rm initial}(x^a) +(\tau-\tau_{\rm a})Y^A_{\rm initial}(x^a) +(\tau-\tau_{\rm a})^2 U^A(\gencoord).
\eea
Note that this formulation, incorporating the initial data, leads us to a form of the eqs.~(\ref{eq:SpecEqGenForm})  which is singular at $\tau=\tau_{\rm a}$ in terms of the fields $U^A(\tau, x^a)$. As a consequence, no initial data are required for $U^A$. We solve this system for $U^A(\gencoord)$ by means of the fully pseudo-spectral  method described in the next section.

%%%%%%%%%%%%%%%%%%%%%%%%%%%%%%%%%%%%%%%%%%%%%%%%%%%%%%%%%
\subsection{(2+1)-Pseudo-Spectral Method}\label{sec:SpecCode}
For prescribed numbers $n_\tau, n_\sigma$ and $n_\theta$ of grid points  in the several directions, we work with Chebyshev Radau collocation points for the time direction $\tau \in [\tau_{\rm a}, \tau_{\rm b}]$ and with Chebyshev Lobatto points in the spatial directions $\sigma \in [\sigma_{\rm a}, \sigma_{\rm b}]$ and $\theta \in [0, \pi]$. In particular, the grid points are given by:
\beq
\label{eq:grid}
\begin{array}{cclcc}
\sigma_i &=& \sigma_{\rm a} + (\sigma_{\rm b}-\sigma_{\rm a})\sin^2\left(\dfrac{\pi i}{2N_\sigma}\right), & i=0...N_\sigma, & N_\sigma=n_\sigma-1, \\
\theta_j &= & \pi \sin^2\left(\dfrac{\pi j}{2N_\theta}\right), &
j=0...N_\theta, & N_\theta=n_\theta-1,\\
\tau_k &= &\tau_{\rm a} +  (\tau_{\rm b} - \tau_{\rm a}) \sin^2\left(\dfrac{\pi}{2}\dfrac{2k+1}{2N_\tau+1}\right),&
k=0...N_\tau, &N_\tau=n_\tau-1.
\end{array}
\eeq
Before describing the pseudo-spectral solution procedure, two comments are worth mentioning:
\ben
\item Note that regarding the angular direction, the spectral decomposition is performed in $\theta$, and {\em not} with respect to $\mu=\cos\theta$, as usually done for axisymmetric problems. Hence, functions  as $f(\theta) =\sin\theta$ are analytic in $\theta$ and have an exponentially converging spectral representation (in contrast to $f(\mu) = \sqrt{1-\mu^2},\, \mu\in[-1,1]$). 
\item Once the solutions $X^A(\gencoord)$ are obtained for $\tau \in [\tau_{\rm a}, \tau_{\rm b}]$, the values $X^A(\tau_{\rm b}, x^a)$ at the upper time boundary $\tau_{\rm b}$ may serve as initial data for a subsequent  time domain $\tau\in [\tau_{\rm b}, \tau_{\rm c}]$. As the upper time bound is included in the grid ($\tau_{k} = \tau_{\rm b}$ for $k=N_\tau$), no need of inter- or extrapolation arises (which could be a source of numerical error). Proceeding in this manner, one can dynamically evolve the system in question until an arbitrary final time value $\tau_{\rm final}$ is reached. Note that with the unsymmetrically distributed Radau collocation points in the time direction, we obtain stable time evolutions for long runs~\cite{Butcher64}.
\een
We carry on by following the common procedure in a pseudo-spectral scheme (see e.g.~\cite{Meinel:2008}), that is, we combine the values 
\[
\Big(X^A(\tau_k,\sigma_i,\theta_j),X^B(\tau_k,\theta_j)\Big)\qquad \mbox{with $A=1,\ldots, n_{\rm fields},\quad B=1,\ldots, n_{\rm boundary}$}
\]
 of the unknown functions at all collocation points (\ref{eq:grid}) to form a vector $\mathbf{X}$. From any such vector, {\em Chebyshev coefficients $c^A_{kij}$} of the fields $X^A=X^A(\coord)$ can be computed by inverting the equations  
\bea
\label{eq:SpecAppr}
X^A(\tau_k,\sigma_i,\theta_j) &=& \sum_{n=0}^{N_\tau}  \sum_{l=0}^{N_\sigma} \sum_{m=0}^{N_\theta} 
c^A_{nlm} T_n\left[\frac{2\tau_k-\tau_{\rm a}-\tau_{\rm b}}{\tau_{\rm b}-\tau_{\rm a}}\right]
          T_l\left[\frac{2\sigma_i-\sigma_{\rm a}-\sigma_{\rm b}}{\sigma_{\rm b}-\sigma_{\rm a}}\right]
          T_m\left[\frac{2}{\pi}\theta_j-1\right].
\eea
Likewise, the Chebyshev coefficients $c^B_{kj}$ for the boundary functions $X^B=X^B(\tau,\theta)$ are obtained by inverting 
\bea
\label{eq:SpecAppr}
X^B(\tau_k,\theta_j) &=& \sum_{n=0}^{N_\tau}  \sum_{m=0}^{N_\theta} 
c^B_{nm} T_n\left[\frac{2\tau_k-\tau_{\rm a}-\tau_{\rm b}}{\tau_{\rm b}-\tau_{\rm a}}\right]
          T_m\left[\frac{2}{\pi}\theta_j-1\right].
\eea
In these equations, $T_\ell$ are Chebyshev polynomials of the first kind, $T_\ell(\xi)=\cos[\ell\arccos(\xi)],\,\xi\in[-1,1]$. 

Now, for the pseudo-spectral solution procedure, we need to compute spectral approximations of first and second time and spatial derivatives of $\{X^A,X^B\}$ at all grid points (\ref{eq:grid}) which we perform by applying specific differentiation matrices to the vector $\mathbf{X}$, see~\cite{canuto_2006_smf}. 
Then, the combination of PDEs and boundary conditions, evaluated at the points (\ref{eq:grid}), yields a non-linear system of algebraic equations of order 
\[n_{\rm total}=n_\tau n_\theta(n_{\rm fields}n_\sigma +n_{\rm boundary})\] for the entries of the vector $\mathbf{X}$.
This system is solved with a Newton-Raphson scheme, which we discuss now in detail.

For a successful application of the Newton-Raphson scheme to solve nonlinear equations, the prescription of an approximate solution of the problem -- a so-called {\em initial-guess} -- is required. If the approximation is not good enough, the iterations inside the scheme diverge and the solution cannot be found. On the other hand, under the weak assumptions that:
\ben[i)]
\item the Jacobian of the function $E^A$ to be zeroed is non-degenerate, and
\item the corresponding second derivatives are finite,
\een
the convergence of the method is always mathematically guaranteed if the initial-guess can be chosen sufficiently close to the solution. In our situation, the initial guess is of the form
\[\mathbf{X}^{(0)}=\Big(X^{A(0)}(\tau_k,\sigma_i,\theta_j),X^{B(0)}(\tau_k,\theta_j)\Big)\qquad\mbox{with $A=1,\ldots, n_{\rm fields},\quad B=1,\ldots, n_{\rm boundary}$},\]
and describes an approximation of the pseudo-spectral space-time solution $\mathbf{X}$ 
of the problem\footnote{Hence, the {\em initial-guess} $\mathbf{X}^{(0)}$ is not to be confused 
with the {\em initial data} $\{X^A_{\rm initial}(x^a), Y^A_{\rm initial}(x^a)\}$ depending only on the spatial coordinates.}.

Now, given an initial-guess $\mathbf{X}^{(0)}$ comprising the quantities $X^{A(0)}$ for $A=1,\ldots,n_{\rm var}$, the scheme approximates the solution iteratively by writing
\beq
X^{A(n+1)} = X^{A(n)} + \delta X^{A(n)}, \nonumber
\eeq
where $\delta X^{A(n)}$ is the solution of the linear system\footnote{\label{footnote:EinsteinConv}We use the Einstein summation convention, according to which in a single term a summation over all the values of an index is implied, if this index variable appears twice in that term.} 
\bea
\displaystyle
\underbrace{\left[ \frac{\partial E^A}{\partial X^B} + \frac{\partial E^A}{\partial X^B_{,\mu}}\frac{\partial}{\partial x^\mu} +  \frac{\partial E^A}{\partial X^B_{,\mu \nu}}\frac{\partial^2}{\partial x^\mu\partial x^\nu}\right]}_{\displaystyle J_B^{A(n)}}\delta X^{B(n)}+ {E}^{A(n)}=0. \label{eq:NR_linSys}
\eea
The functions  ${E}^{A(n)}$ (describing inhomogeneous or source terms) as well as the Jacobian $J_B^{A(n)}$ depend on time and on the $n$-th iterative solution $X^{A(n)}$, i.e.,  
\bea
\nn		{E}^{A(n)} &=& {E}^A\left(\tau, \left\{{X}^{B}{}^{(n)}\right\}, \left\{\partial_\tau {X}^{B(n)}\right\}, \left\{\partial^2_{\tau\tau} {X}^{B(n)}\right\} \right),\\
 \nn     J_B^{A(n)} &=& {J}_B^{A(n)}\left(\tau, \left\{{X}^{B}{}^{(n)}\right\}, \left\{\partial_\tau {X}^{B(n)}\right\}, \left\{\partial^2_{\tau\tau} {X}^{B(n)}\right\} \right).
\eea

Note that the equations (\ref{eq:NR_linSys}) look like a system of linear PDEs for the unknown variable  $\delta X^A$. However, one should think of the differentiations appearing in (\ref{eq:NR_linSys}) as the application of corresponding spectral differentiation matrices by which, for finite resolution $n_{\rm total}$, a discrete linear problem is rendered. In the continuum limit of infinite resolution, the equations (\ref{eq:NR_linSys}) pass into a system of linear PDEs.

In the following, we shall express (\ref{eq:NR_linSys}) with the notation

\beq
L^A\left(\tau, \left\{\delta X^{B(n)}\right\}, \left\{\partial_\tau {\delta X}^{B(n)}\right\}, \left\{\partial^2_{\tau\tau} {\delta X}^{B(n)}\right\}; X^{B(n)}, E^{B(n)}\right) = 0, \label{eq:NR_linSys2}
\eeq
suppressing the fact that $L^ A$ depends also on first and second time and spatial derivatives of $X^{B(n)}$. Numerically, the problem of solving the linear system (\ref{eq:NR_linSys}) is equivalent to inverting the dense matrix $J_B^{A(n)}$. 

\vspace*{1mm}
\noindent
At this point we have to address the following two questions:  
\ben
	\item How can we efficiently invert $J_B^{A(n)}$ and solve the linear system (\ref{eq:NR_linSys}) for ${\delta X}^{B(n)}$ ?
	\item How can we obtain a sufficiently good initial-guess $\mathbf{X}^{(0)}$?
\een
For typical values of $N_\tau, N_\sigma$ and $N_\theta$, inverting  $J_B^{A(n)}$ by a direct method (such as LU decomposition) is an extremely expensive approach and unfeasible in many situations.  Therefore, we go about the application of an iterative method, specifically the bi-conjugate gradient stabilized method ("BiCGStab", see e.g.~\cite{Barrett93}). Now, in order to obtain a reasonably small number of iterations in that scheme, we need to provide the BiCGStab method with a pre-conditioner. Here we utilize an approximative  solution of eq.~(\ref{eq:NR_linSys}) which we compute with the help of a singly diagonally implicit Runge-Kutta method ("SDIRK", see~\cite{butcher2008numerical,Alexander77,Skvortsov06}).

Appendix \ref{app:BiCGStab} presents the BiCGStab algorithm adapted to our notation. Note that, at a given Newton-Raphson iteration $n$, the BiCGStab method requires the approximate solution of multiple linear systems of the type 
\beq
\label{eq:BiCGstab_linSys}
L^A\left(\left\{ W^B\right\}, \left\{\partial_\tau W^B\right\},\left\{\partial^2_{\tau\tau} W^B\right\}; X^{B(n)}, b^B\right)=0,
\eeq
for various source terms $b^A$ (which differ inside the BiCGStab solution procedure from $E^{A(n)}$, see eq. (\ref{eq:NR_linSys2})). Hereby, we obtain the approximate solution ${W}^A(\tau,x^a)$ with the SDIRK method which we present in detail in section \ref{sec:SDIRK}. 

We apply the SDIRK method another time, as we use it to solve approximately the nonlinear system (\ref{eq:SpecEqGenForm}). In this manner we compute our initial-guess  $\mathbf{X}^{(0)}$ for the Newton-Raphson scheme. 

%%%%%%%%%%%%%%%%%%%%%%%%%%%%%%%%%%%%%%%%%%%%%%%%%%%%%%%%%
\subsection{Singly Diagonally Implicit Runge Kutta (SDIRK-) method}\label{sec:SDIRK}
In our applications, the SDIRK method aims at the solution of a system of equations written implicitly in the generic compact form
\beq
\label{eq:RKgenForm}
F^A\left(\tau, \left\{ X^B\right\}, \left\{ \frac{\partial X^B}{\partial\tau} \right\}, \left\{ \frac{\partial^2 X^B}{\partial\tau^2}\right\} \right) = 0.
\eeq
As described above, our usage of the SDIRK method is twofold. On the one hand, the SDIRK serves as a pre-conditioner inside the BiCGStab method used to iteratively solve the linear problem (\ref{eq:NR_linSys}). In this situation, $F^A$ represents the system described in (\ref{eq:BiCGstab_linSys}), with $W^A$ as unknown variable. In the second application, the SDIRK method provides us with an initial-guess for the Newton-Raphson scheme. For this purpose we choose $F^A$ to be completely equivalent to $E^A$ introduced in eq. (\ref{eq:SpecEqGenForm}).  

At this point it is essential to assume that the PDE system in question be {\em quasi-linear}\footnote{We call a PDE 
of second order  {\em quasi-linear} if it is linear in the second order derivatives of the unknown functions, with
coefficients that may depend non-linearly on the independent variables as well as on the unknown functions and their  first derivatives.}. In other words, we require $F^A$ to be linear in its last argument $\left\{ \dfrac{\partial^2 X^B}{\partial\tau^2}\right\}$. Note that it is desirable to write the boundary conditions $F^A$ for $A>n_{\rm fields}$ also in a quasi-linear second order form. In cases in which the boundary conditions $F^A$ are of lower order (such as e.g.~conditions (\ref{eq:WaveEqBoundCond})), a quasi-linear second order form is obtained by differentiating the conditions with respect to the time $\tau$. In this way, the conditions $E^2$ and $E^3$ arise in our example of the free boundary hyperbolic problem of the wave equation, see (\ref{eq:WaveEqFreeBound_F2}, \ref{eq:WaveEqFreeBound_F3}).

We now introduce the notion of singly diagonally implicit methods. To this end, we review the general scheme of Runge-Kutta methods of order $s$ for ordinary differential equations (ODEs) of first order. Given a system of equations of the form 
\beq
\label{eq:RKForm}
\dot X^A  = f^A(\tau, X^B), \qquad\mbox{where}\quad \dot{}\equiv \frac{d}{d\tau},
\eeq
together with initial values $X^A(\tau_{\rm a})=X^A_{\rm initial}$, the time integration (with $X^A(\tau_k)=X^A_k$) is performed by writing
\bea
X^A_{k+1} &=& X^A_k +h \sum_{(i)=1}^s b_{(i)}K^A_{(i)}, \quad {\rm with} \label{eq:RKevol} \\
K^A_{(i)} &=& f^A \left( \tau_k + hc_{(i)}, X^B_k  +  h\sum_{(j)=1}^s a_{(i)(j)}K^B_{(j)} \right).\label{eq:K}
\eea
Here $(i), (j), ...$ are indexes inside the Runge-Kutta method that run from 1 to the order number $s$. The coefficients $a_{(i)(j)}$, $b_{(j)}$ and $c_{(j)}$ are determined by a Butcher tableau~\cite{butcher2008numerical}, specifically
\beq
\label{tab:SDIRKtableau}
\left.
\begin{array}{c|cccc}
  c_1    & a_{11} & a_{12}& \dots & a_{1s}\\
  c_2    & a_{21} & a_{22}& \dots & a_{2s}\\
  \vdots & \vdots & \vdots& \ddots& \vdots\\
  c_s    & a_{s1} & a_{s2}& \dots & a_{ss}\\
  \hline & b_1    & b_2   & \dots & b_s
    \end{array}
    \right\}
\begin{array}{l}
\mbox{General}\\ \mbox{Runge-Kutta}\\ \mbox{method}
\end{array}
\qquad\quad
\left.
\begin{array}{c|cccc}
  c_1    & \gamma & 0 & \dots & 0\\
  c_2    & a_{21} & \gamma& \dots & 0\\
  \vdots & \vdots & \vdots& \ddots& \vdots\\
  c_s    & a_{s1} & a_{s2}& \dots & \gamma \\
  \hline & b_1 & b_2   & \dots & b_s
\end{array} 
    \right\}
\begin{array}{l}
\mbox{Singly-Diagonally}\\ \mbox{Implicit Runge-Kutta}\\ \mbox{method (SDIRK)}
\end{array}
\eeq

An explicit scheme is characterized by $a_{(i)(j)}=0$ for $(j)\ge (i)$, while an implicit method has $a_{(i)(j)} \ne 0$ for $(j)\ge(i)$. In implicit methods, eqs.~(\ref{eq:K}) represent for an ODE a system of algebraic equations for the $K^A_{(i)}$, whereas for PDEs in (1+$d$)-dimensions, they correspond to a system of spatial differential equations in $d$ dimensions. By restricting ourselves to Diagonally Implicit Runge-Kutta methods (DIRK), in which $a_{(i)(j)} = 0$ for $(j)>(i)$, the system decouples with respect to the Runge-Kutta index $(i)$, i.e., for each $(i)=1...s$, eq.~(\ref{eq:K}) describes implicit equations for the unknowns $K^A_{(i)}$ only. In other words, when solving iteratively for each $K^A_{(i)}$, all the previously calculated $K^A_{(j)}$ with $(j)<(i)$ enter the equation as source terms. Finally, Singly Diagonally Implicit Runge-Kutta methods (SDIRK) possess a constant diagonal term $a_{(i)(i)}=\gamma$. 

In this work, we use the following $s=3$ stage Butcher tableau
\bea
\label{eq:usedtableau}
\begin{array}{c|ccc}
  \gamma    & \gamma & 0 &  0\\
  (1+\gamma)/2    & (1-\gamma)/2 & \gamma&  0\\
  1    & 1-b_2-\gamma & b_{2} & \gamma \\
  \hline & 1-b_2-\gamma & b_2   & \gamma
\end{array}
\eea
with $b_2 = (5-20\gamma + 6\gamma^2)/4$ and\footnote{According to \cite{Skvortsov06}, $\gamma$ is a root of the polynomial $\frac{1}{6}-\frac{3}{2}\gamma+3\gamma^2-\gamma^3$.} $\gamma = 0.435866521508$. This scheme was suggested in \cite{Alexander77}; it has convergence order $3$ and is widely used in problems dealing with stiff ODEs (see also \cite{Skvortsov06}). The detailed description of the corresponding stability properties is beyond our objective (see instead \cite{butcher2008numerical,Alexander77,Skvortsov06}). It is, however, important to note that the tableau ($\ref{eq:usedtableau}$) is classified as L-Stable, which, for the practical purpose we are interested in, means that the stability of the solution during the time integration is not affected by the size of the time step $h$.

In order to adapt the SDIRK method presented above for ODEs to our system of PDEs (\ref{eq:RKgenForm}), we need to address the question how implicitly given equations can be treated with a Runge-Kutta method which uses the explicit form (\ref{eq:RKForm}). We answer this question by writing (\ref{eq:RKForm}) in the form
\beq
\label{eq:RKForm_implicit}
F^A\left(\tau, \left\{ X^B\right\}, \left\{ \dot X^B \right\} \right) := 
f^A(\tau, X^B)-\dot X^A  = 0,
\eeq
and, furthermore, eqn.~(\ref{eq:K}) as:
\beq
\label{eq:K_implicit}
F^A\left(\tau_k + hc_{(i)}, \left\{ X^B_k  +  h\sum_{(j)=1}^s a_{(i)(j)}K^B_{(j)}\right\}, 
\left\{ K^B_{(i)} \right\} \right) = 0.
\eeq

For explicit equations, the formulations (\ref{eq:RKForm}) and (\ref{eq:K}) are completely equivalent to (\ref{eq:RKForm_implicit}) and (\ref{eq:K_implicit}). We now assume that the latter formulation can be carried over to the realm of general implicit ODEs of the form
\[
F^A\left(\tau, \left\{ X^B\right\}, \left\{ \dot X^B \right\} \right) =0,
\]
and that the solution via a Runge-Kutta method at the time $\tau_{k+1}$ is given through eqn.~(\ref{eq:RKevol}) where the quantities $K^A_{(i)}$ are computed from (\ref{eq:K_implicit}) which is a system of implicit algebraic equations. This assumptions is strongly confirmed through several numerical experiments.

Another point to be clarified before returning to the system of PDEs regards the reduction of second order ODEs to a system of first order. This step is necessary since Runge-Kutta schemes are tailored to the treatment of first order ODEs. Given a system of implicit second order ODEs,
\[
F^A\left(\tau, \left\{ X^B\right\}, \left\{ \dot X^B \right\}, \left\{ \ddot X^B \right\} \right) =0,
\]
we introduce the auxiliary variable
\[Y^A=\dot X^A\]
and obtain the first order system
\bea
\label{eq:vec_F^A}
\vec{F}^A\left(\tau,\left\{\vec X^B\right\}, \left\{\dot {\vec X}^B\right\}\right)
& \equiv &\left( \begin{array} {c}  Y^A - \dot X^A\\ F^A\left(\tau, \left\{ X^B\right\}, \left\{ Y^B \right\}, \left\{ \dot Y^B\right\} \right) \end{array}\right)=0,
\eea
where we have written
\[
\vec{X}^A=\left( \begin{array} {c}  X^A \\ Y^A \end{array}\right) \nonumber .
\]
As described above, the Runge-Kutta method yields approximate solutions $\vec X^A_{k+1}$ to the system (\ref{eq:vec_F^A}) at the time $\tau_{k+1}$ via eqns.~(\ref{eq:RKevol}) and (\ref{eq:K_implicit}), now written in terms of two-dimensional vectorial quantities $\vec{X}^A, \vec{F}^A$ and $\vec K_{(i)}^A$,
\bea
\vec X^A_{k+1} &=& \vec X^A_k +h \sum_{(i)=1}^s b_{(i)}\vec K^A_{(i)}, \quad {\rm with} \label{eq:vecRKevol}\\
\label{eq:vecK_implicit}
0&=&\vec F^A\left(\tau_k + hc_{(i)}, \left\{ \vec X^B_k  +  h\sum_{(j)=1}^s a_{(i)(j)}\vec K^B_{(j)}\right\}, 
\left\{\vec  K^B_{(i)} \right\} \right).
\eea
If we define for an SDIRK scheme (\ref{tab:SDIRKtableau})
\beq\label{eq:SDIRK_M}
\vec{M}^A_{(i)}  :=\left\{\begin{array}{ll}
\vec{X}^A_k & \mbox{for $(i)=1$,} \\[3mm]
\vec{X}^A_k +h {\displaystyle\sum\limits_{(j)=1}^{(i)-1} a_{(i)(j)}\vec{K}^A_{(j)}}
& \mbox{for $(i)>1$} 
\end{array}\right.
\eeq
and write, furthermore, 
\beq\label{eq:SDIRK_MK}
\vec{M}^A_{(i)}=\left( \begin{array} {c}  M^A_{(i)} \\[3mm] N^A_{(i)} \end{array}\right), \qquad 
\vec{K}^A_{(i)}=\left( \begin{array} {c}  K^A_{(i)} \\[3mm] G^A_{(i)} \end{array}\right),
\eeq
we obtain the following explicit expression
\beq
\nn
\vec{F}^A\left(\tau,
\left\{\left(\begin{array} {c}  X^B \\ Y^B \end{array} \right)\right\}, 
\left\{\left(\begin{array} {c}  \dot X^B \\ \dot Y^B \end{array} \right)\right\}
\right)
=\left( 
\begin{array} {c}  
		Y^A - \dot X^A\\ 
		F^A\left(\tau, \left\{ X^B\right\}, \left\{ Y^B \right\}, \left\{ \dot Y^B \right\}\right)
\end{array}\right)
\eeq
for eqn.~(\ref{eq:vec_F^A}) and
 \bea
 \lefteqn{\nn
\label{eq:KG_implicit}
\vec{F}^A\left(\tau_k + hc_{(i)},
\left\{\left(\begin{array} {c}  M^B_{(i)}+h\gamma K_{(i)}^B \\[2mm] N^B_{(i)}+h\gamma G_{(i)}^B \end{array} \right)\right\}, 
\left\{\left(\begin{array} {c}  K_{(i)}^B \\[2mm] G_{(i)}^B \end{array} \right)\right\}
\right)}&&\\[3mm] 
&=&\left( 
\begin{array} {c}  
		N_{(i)}^A+h\gamma G_{(i)}^A - K_{(i)}^A\\[2mm]
		F^A\left(\tau_k + hc_{(i)}, \left\{ M^B_{(i)}+h\gamma K_{(i)}^B\right\}, 
		                            \left\{ N^B_{(i)}+h\gamma G_{(i)}^B \right\}, 
		                            \left\{ G_{(i)}^B \right\}\right)
\end{array}\right)=0
\eea
for eqn.~(\ref{eq:vecK_implicit}). The first component immediately gives
\beq\label{eq:SDIRK_K}
	K_{(i)}^A=N_{(i)}^A+h\gamma G_{(i)}^A,	
\eeq
which we may insert into the second component, thus yielding
\beq\label{eq:SDIRK_G_implicit}
	F^A\left(\tau_k + hc_{(i)}, \left\{ M^B_{(i)}+h\gamma N_{(i)}^B+h^2\gamma^2 G_{(i)}^B\right\}, 
		                            \left\{ N^B_{(i)}+h\gamma G_{(i)}^B \right\}, 
		                            \left\{ G_{(i)}^B \right\}\right)=0.
\eeq
For second order ODEs, (\ref{eq:SDIRK_G_implicit}) is a set of nonlinear implicit 
algebraic equations for the quantities $G_{(i)}^A$. Beginning with $ (i)=1$, this set is to be solved for all Runge-Kutta stages $ (i)=1,\ldots s $. At a specific stage $(i)$, the quantities $M_{(i)}^B$ and $N_{(i)}^B$ are known from the data
$\vec X^B_k$ at the point $\tau_k$ and the results of previous stages, cf.~(\ref{eq:SDIRK_M}) and (\ref{eq:SDIRK_MK}). Note that, in the limit of vanishing time step size $ h $, the dependency of $ F^A $ on $G_{(i)}^B$ only survives in its last argument. Now, as described at the beginning of subsection \ref{sec:SDIRK}, this dependency is assumed to be linear. Hence, for sufficiently small step sizes $h$, the system 
(\ref{eq:SDIRK_G_implicit}) is {\em almost} linear, and a Newton-Raphson-method with the trivial initial guess $G_{(i)}^A=0$ converges.
Once $G_{(i)}^A$ has been found, we compute via (\ref{eq:SDIRK_K}) and (\ref{eq:SDIRK_MK}) the vector $\vec K_{(i)}^A$ which is not only necessary to perform the entire Runge-Kutta step (\ref{eq:vecK_implicit}) but also for the computation of $\vec M_{(i)+1}^B$ used in the next Runge-Kutta stage, cf.~(\ref{eq:SDIRK_M}).

At this point we return to our original objective, the treatment of hyperbolic PDEs. Thanks to the convention made at the beginning of section \ref{sec:NumMeth} for the notation $\{\bullet\}$ as involving all kinds of spatial derivatives, we arrive at the same set of equations (\ref{eq:SDIRK_G_implicit}). Now, however, this system comprises nonlinear differential equations in spatial dimensions for the functions $G^A_{(i)}$.  For our generic (2+1)-problem, with $n_{\rm fields}$ field variables depending on the space-time coordinates $(\tau, \sigma, \theta)$ and $n_{\rm boundary}$ boundary variables depending on $(\tau, \theta)$, the set  (\ref{eq:SDIRK_G_implicit}) consists of spatial PDEs in   $(\sigma, \theta)$ for $G_{(i)}^A$ corresponding to field variables and ODEs in $\theta$ for boundary variables. For each Runge-Kutta stage $(i)$, the equations (\ref{eq:SDIRK_G_implicit}) are to be solved {\em simultaneously} in order to obtain the entirety of quantities $G^A_{(i)}$ for $A=1,\ldots,n_{\rm var}= n_{\rm fields}+n_{\rm boundary}$. Again we note that for sufficiently small step sizes $h$, the system is nearly linear through the requirement of quasi-linearity of our problem, and hence a solution can be found utilizing a Newton-Raphson-method with the trivial initial guess $G_{(i)}^A=0$. The subsequent steps are exactly the same as described above for ODEs. The knowledge of $G_{(i)}^A$ implies that of $\vec K_{(i)}^A$ and $\vec M_{(i)+1}^B$ by means of which the next Runge-Kutta stage as well as eventually the entire Runge-Kutta step can be done.

Depending on the specific situation, the system (\ref{eq:SDIRK_G_implicit}) can comprise singular elliptic PDEs to be solved without an additional requirement of boundary conditions. In other situations, the boundary terms appear as unknown parameters in the PDE and the system is closed by virtue of supplemental boundary conditions. In any case, the set of spatial PDEs and boundary conditions is inherited from the original (possibly free boundary) hyperbolic problem of second order PDEs. 

The situation can be better appreciated with the example introduced previously at the beginning of section \ref{sec:NumMeth}  which represents a specific free boundary hyperbolic problem of the (1+1)-wave equation. For convenience, we reproduce here\footnote{As we want to treat eqs. (\ref{eq:WaveEqFreeBound_F1})-(\ref{eq:WaveEqFreeBound_F3}) within the context of the SDIRK Runge-Kutta method,  the equations are here referred to as $F^A$ instead of $E^A$. See discussion at the beginning of section \ref{sec:SDIRK}.} eqs.~(\ref{eq:WaveEqFreeBound_F1})-(\ref{eq:WaveEqFreeBound_F3})

\bea
F^1\left(\tau, \{X^A\}, \{Y^A\}, \{Z^A\}\right) &=& \left(X^2-X^3\right)^2Z^1 - 
\left\{4T_0^2- \left[(1-\sigma)Y^3 + (1+\sigma)Y^2\right]^2\right\}\left(X^1\right)''\nonumber \\
&& + 2\left[(1-\sigma)Y^3 + (1+\sigma)Y^2 \right] \left[ \left(X^1\right)'\left( Y^2 - Y^3 \right) - \left(Y^1\right)'(X^2-X^3)\right]\nonumber \\
&&-(X^2-X^3)\left[(1-\sigma)Z^3 + (1+\sigma)Z^2\right](X^1)'  \nn \\
F^2\left(\tau, \{X^A\}, \{Y^A\}, \{Z^A\}\right) &=& \left. Z^1\right|_{\sigma = -1} - \ddot \phi^-(\tau),\nn \\
F^3\left(\tau, \{X^A\}, \{Y^A\}, \{Z^A\}\right) &=& \left. Z^1\right|_{\sigma = +1} - \ddot \phi^+(\tau). \nn
\eea 

As depicted above, the system (\ref{eq:WaveEqFreeBound_F1})-(\ref{eq:WaveEqFreeBound_F3}) needs to be cast into the form (\ref{eq:SDIRK_G_implicit}), in which (\ref{eq:WaveEqFreeBound_F1}) describes a spatial second order  ODE in $\sigma$ for $G^1_{(i)}$ that contains unknown parameters $G^2_{(i)}$ and $G^3_{(i)}$. Then, the entire system is uniquely soluble by virtue of the supplemental boundary conditions (\ref{eq:WaveEqFreeBound_F2}) and (\ref{eq:WaveEqFreeBound_F3}). 

Note that for the unique solution of this ODE no further boundary conditions are needed. The reason is that, for  $Y^3\geq T_0$ and $Y^2\leq-T_0$, the coefficient $C(\tau,\sigma) $ in front of $\left( X^1\right)''$ vanishes two times in $[-1,1]$. In fact, the coefficient can be written as 
\[C(\tau,\sigma)=A^+(\tau,\sigma)A^-(\tau,\sigma) \quad {\rm with} \quad
A^\pm(\tau,\sigma)=(1-\sigma)Y^3 +(1+\sigma)Y^2 \pm 2T_0,
\]
whence
\[A^\pm(\tau,-1) = 2(Y^3 \pm T_0) \geq 0,\qquad A^\pm(\tau,1) = 2(Y^2 \pm T_0) \leq 0.\] 

\begin{figure*}[h!]
\begin{center}
\includegraphics[width=9.cm]{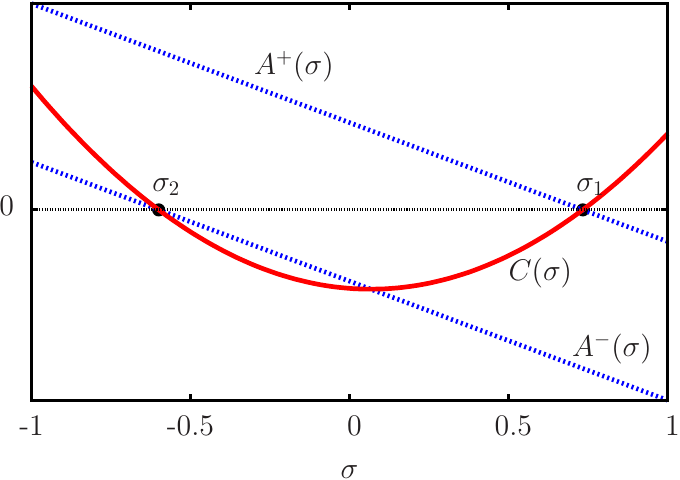}
\end{center}
\caption{Roots of the function $C(\sigma)=A^+(\sigma)A^-(\sigma)$. They correspond to the points where the spatial differential equation (\ref{eq:WaveEqFreeBound_F1}) degenerates.
 }
\label{fig.FuncDeg}
\end{figure*}

The behavior of $A^\pm$ as well as that of $C$ is illustrated in fig.~\ref{fig.FuncDeg}. As both functions $A^\pm$ are linear in $\sigma$ and have their zero in $[-1,1]$, the coefficient $C=A^+A^-$ possesses exactly two roots $\sigma_{1/2}$ at which the ODE degenerates, giving there a relation involving only up to first order derivatives\footnote{Note that for $T_0>0$ we have $A^+\neq A^-$, and hence the two roots differ always, $\sigma_1\ne \sigma_2$.}. Now, the specific character of our ODE implies that by virtue of these degeneracies {\em only one} regular solution can be found. This reflects the fact that the characteristics of our original wave equation always point outward the numerical domain, provided that ${dx^-}/{d\tau}\geq T_0$ and ${dx^+}/{d\tau}\leq-T_0$, and hence, no further boundary conditions are required.
We find an analogous behavior in more complicated systems of equations.

We finish this section by commenting that spatial eqs.~(\ref{eq:SDIRK_G_implicit}) are solved numerically again with a pseudo-spectral method, as the one described in \cite{Meinel:2008}. In general terms, the solution is obtained following the ideas presented in section \ref{sec:SpecCode}. For (2+1)-problems, the matrices appearing in the context of the spatial differential equations are inverted once more with the iterative BICGStab method. Here, the pre-conditioner relies on the inversion of a band matrix which appears as a finite-difference approximation to the Jacobian matrix, see \cite{Meinel:2008} for details. We note that this spatial solver always converges even for  coarse resolutions.

%%%%%%%%%%%%%%%%%%%%%%%%%%%%%%%%%%%%%%%%%%%%%%%%%%%%%%%%%
\subsection{An example: the (1+1)-wave equation with free boundaries}
We conclude this section with the presentation of a specific numerical solution to the example introduced at the beginning of  section \ref{sec:NumMeth}. Even though the system of equations (\ref{eq:WaveEqFreeBound_F1})-(\ref{eq:WaveEqFreeBound_F3}) is just a different representation of the simple wave equation $-\phi_{,tt} + \phi_{,xx}=0$, it includes all the complications which we will have to deal with when solving other problems, namely, the presence of {\em unknown free boundaries} (treated in terms of the coordinate transformation (\ref{eq:WaveEqCoordTrans})), the {\em nonlinearity} (coming from the coupling between $X^1$ with $X^2$ and $X^3$) and the occurrence of {\em implicit} equations. Such a set-up goes beyond the framework in which the SDIRK method is usually applied (as mentioned, tableau (\ref{eq:usedtableau}) was developed for {\em explicit} ODEs and optimized for problems with stiff terms), and we check here applicability and efficiency of the extended SDIRK method.

\begin{figure*}[h!]
\begin{center}
\includegraphics[width=8.2cm]{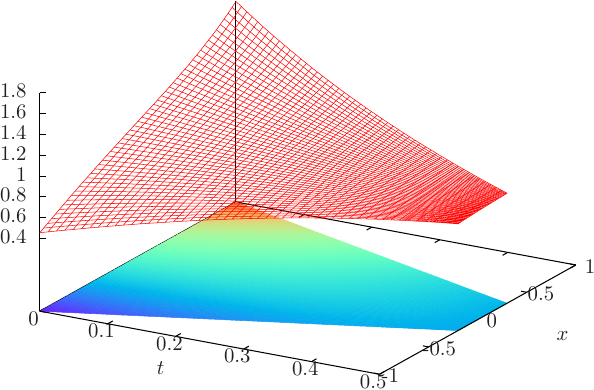}
\includegraphics[width=8.1cm]{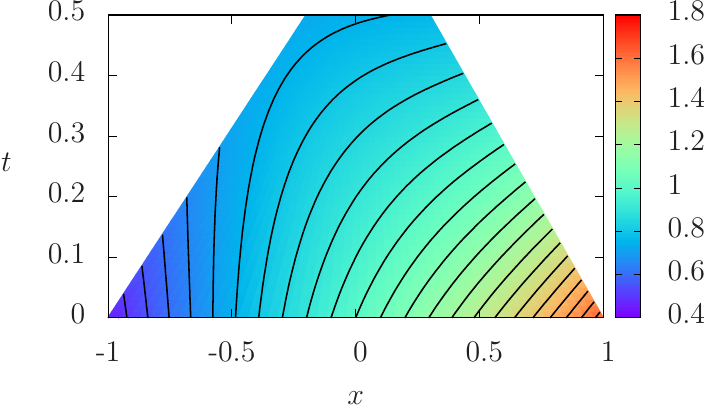}
\end{center}
\caption{Plots of the explicitly prescribed solution (\ref{eq:WaveEqFreeBound_sample}) (see also (\ref{eq:phi_exact}) and (\ref{eq:WaveEqCoordTrans})) to the free boundary problem of the (1+1)-wave equation.
 }\label{fig:WaveEqFreeBound}
\end{figure*}

As a specific example we take:
\bea
f(t+x) &=& \cos(t+x), \nonumber \\
g(t-x) &=& e^{t-x}, \label{eq:WaveEqFreeBound_sample} \\
x^\pm_{\rm exact}(\tau)&=& \pm \left[ 1 - \left(T_0  +\epsilon^\pm\right) \tau  \right] \quad {\rm with} \quad \epsilon^+=0.2 \quad {\rm and} \quad\epsilon^-=0.3. \nonumber
\eea
We insert these expressions into (\ref{eq:phi_exact}) (see also (\ref{eq:WaveEqCoordTrans})) and obtain an explicitly known solution. Corresponding plots can be found in fig.~\ref{fig:WaveEqFreeBound}. Now, in the form of (\ref{eq:phi_exact_ID}), we read off boundary and initial data and numerically solve the corresponding free boundary value problem to determine $\phi(\tau,\sigma)$ and $x^\pm(\tau)$ for $(\tau,\sigma)\in[0,1]\times[-1,1]$. \footnote{The spectral method provides us with approximate values at the discrete spectral grid points (\ref{eq:grid}). Via standard interpolation schemes based on Clenshaw's algorithm \cite{Press92}, approximate values are accessible at any $(\tau,\sigma)\in[0,1]\times[-1,1]$.} 

\begin{figure*}[h!]
\begin{center}
\includegraphics[width=8.1cm]{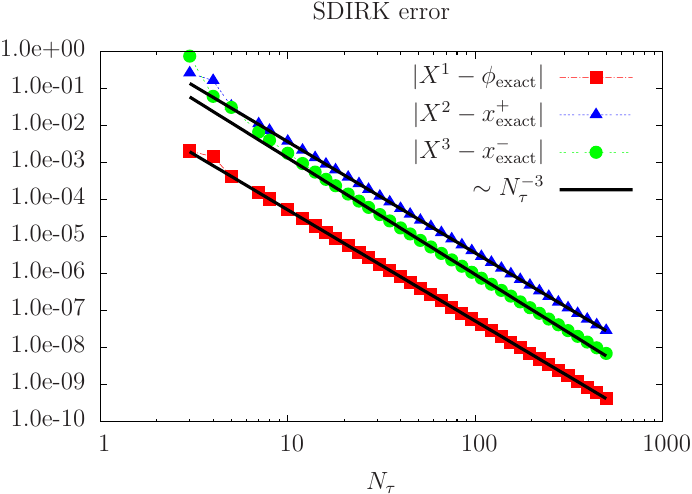}
\includegraphics[width=8.1cm]{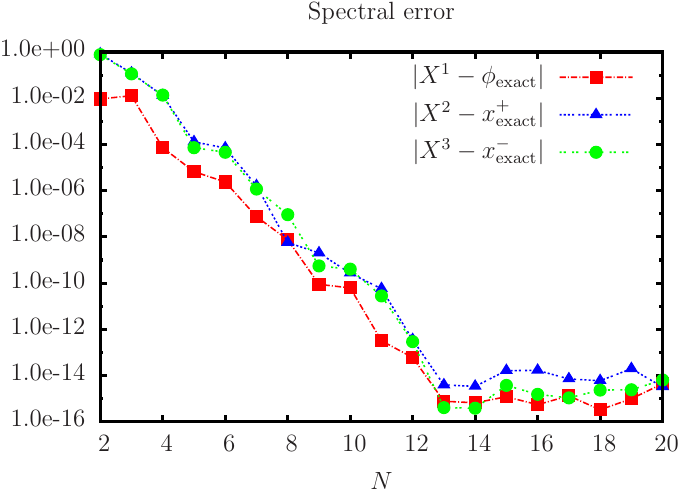}
\end{center}
\caption{Convergence plot for the field variables $X^1(\tau,\sigma)$ (red squares), $X^2(\tau)$ (blue triangles)  and $X^3(\tau)$ (green circles). We obtain the third order convergence expected from the SDIRK method (left panel) and the exponential convergence with the fully spectral method (right panel). The values were taken at $\tau = 1$ ($t=T_0$). As a representative measure of the error, $X^1$ was evaluated at $\sigma=0.5$. Note that the SDIRK integrator might not converge for very coarse resolutions $N_\tau \lesssim 5$
 }
\label{fig.WaveEqFreeBound_ConvPlot}
\end{figure*}

Fig.~\ref{fig.WaveEqFreeBound_ConvPlot} shows the convergence plot for the numerical solutions $\phi, x^\pm$ against $\phi_{\rm exact}$ and $x^\pm_{\rm exact}$. The parameter $T_0$ was chosen to be $T_0=0.5$. As expected for the third order SDIRK method used to find the initial guess for the Newton-Raphson scheme inside the pseudo-spectral solver, the absolute error for the three field variables decreases with  increasing time resolution $N_\tau$ at third order (here we specified a fixed spatial resolution of $N_\sigma = 30$). The initial guess turned out to be sufficiently good (for all time resolutions considered), as the subsequent Newton-Raphson scheme converges rapidly within a few steps. Note that fig.~\ref{fig.WaveEqFreeBound_ConvPlot} also shows the exponential convergence of the fully spectral method against the resolution  $N_\sigma=N_\tau = N$. At a resolution around $N = 15$ the spectral solution is already correct up to $\sim 14$ digits, while the SDIRK method requires $N_\tau > 1000$ to reach machine precision.

%%%%%%%%%%%%%%%%%%%%%%%%%%%%%%%%%%%%%%%%%%%%%%%%%%%%%%%%%
\section{Applications}\label{sec:Applications}
This section is devoted to the application of the fully spectral method to some physical problems. Since our algorithm is a further development of the one presented in~\cite{Hennig:2012zx}, we follow the steps presented therein and solve at first the equations describing spherically symmetric perturbations of a polytropic star within Newton's theory of gravity. Our objective is to show that our method can handle much stronger perturbations than the ones considered in \cite{Hennig:2012zx}. This follows from the fact that our algorithm computes the initial guess for the Newton-Raphson scheme via the SDIRK-method, i.e.~no knowledge of an approximate solution is required. Moreover, we demonstrate that the iterative BiCGstab method equipped with the SDIRK pre-conditioner allows us to solve the equations more efficiently.

The second application is the treatment of axisymmetric problems, specifically the so-called {\em Teukolsky equation}~\cite{Teukolsky73}, which operates in the linear regime of General Relativity and  describes scalar, magnetic and gravitational perturbations of rotating black holes. We consider the formulation of this equation within the conformal context, i.e.~on compactified hyperboloidal slices. In so doing, we obtain a situation in which at the finite numerical boundaries the equation degenerates. As a consequence, the characteristics of the equation do not point inwards, which in turn means that no boundary conditions are required. This set-up provides a good starting point for the study of the fully spectral method in (2+1)-dimensions. In particular, we can compare our numerical findings with some results  available in the literature~\cite{Racz:2011qu, Harms:2013ib}. 

\subsection{Spherically symmetric perturbations of Newtonian stars}
\subsubsection{The physical model}
Our first application is a specific free boundary problem modeling the dynamics of a neutron star within Newton's theory of gravity. Here we follow~\cite{Hennig:2012zx} and consider the star as a polytropic perfect fluid ball with an unknown time-dependent radius $R$. The polytropic exponent $\Gamma$ is chosen to be $\Gamma=2$. The field variables are the gravitational potential $U$, the mass density $\rho$ (which is related to the fluid pressure $P$ according to the polytropic equation of state) and the velocity field $v = r\cdot w$ where $r$ denotes the radial distance coordinate. We pass to dimensionless quantities
\bea
r\rightarrow r_0r, \quad t\rightarrow \frac{r_0}{\sqrt{K\rho_c}}t, \quad
U\rightarrow K\rho_cU, \quad \rho\rightarrow \rho_c\rho, \quad w\rightarrow \frac{\sqrt{K\rho_c}}{r_0}w,\qquad
\mbox{with}\quad r_0=\sqrt{\frac{\pi G}{2K}}, \label{eq:dimensionless}
\eea
where $G$ is the gravitational constant, $K$ the polytropic constant, $\rho_c$ and $r_0$ the central pressure and stellar radius, respectively, of a reference equilibrium configuration. As usual, $t$ denotes coordinate time.

In a first step, coordinates $(\tau, \sigma)\in[0,1]\times[0,1]$ are introduced, such that the pseudo-spectral representations of the fields $U$, $\rho$, $w$ and the boundary $R$ are considered on the grid spanned by $\tau$ and $\sigma$. Specifically, we write
\beq
t=t_n +  \tau\, \Delta t ,\qquad r=\sqrt{\sigma}R(\tau),
\eeq
where we have divided the time interval $[t_{\rm min}, t_{\rm max}]$ into $n_{\rm max}$ sub-intervals 
$[t_0,t_1], [t_1,t_2],...,[t_{n_{\rm max}-1},t_{n_{\rm max}}]$ with 
\[ t_n=t_{\rm min}+n\Delta t, \qquad\Delta t=(t_{\rm max}-t_{\rm min})/n_{\rm max}.\]
The relation $r\leftrightarrow\sigma$ in terms of the unknown function $R$ guarantees that the numerical boundary $\sigma=1$ always describes the surface of the star. Now, we consider the problem on each time interval $[t_n,t_{n+1}]$ separately and take the computed solution at $t=t_{n+1}$ (i.e.~$\tau=1$) as initial data for the subsequent time interval $[t_{n+1},t_{n+2}]$.

The details regarding the derivation of the equations in spherical symmetry are well explained~\cite{Hennig:2012zx}, and we merely list them here in terms of the coordinates $\{\tau, \sigma\}$:
\bea
&&\left\{ 
\begin{array} {ccccl}  
2\sigma U_{,\sigma\sigma} + 3U_{,\sigma} &=& \pi^2 R^2\rho &\text{for}  &\sigma\in[0,1) \\[2mm]
2\sigma U_{,\sigma} + U                  &=& 0             & \text{for} & \sigma=1
 \end{array}\right. \label{eq:U}\\[2mm]
&& R\frac{w_{,\tau}}{\Delta t} + 2\sigma\left( R w - \dfrac{R_{,\tau}}{\Delta t} \right)w_{,\sigma} + Rw^2 + \frac{2}{R}\left( 2\rho_{,\sigma} +  U_{,\sigma}\right) = 0 \label{eq:w} \\[2mm]
&&\frac{\rho_{,\tau}}{\Delta t} + 2\sigma\left( w- \frac{R_{,\tau}}{R\Delta t}\right)\rho_{,\sigma} + \rho\left( 2\sigma w_{,\sigma} + 3 w\right) = 0 \label{eq:rho} \\[2mm]
&&\rho(\tau, \sigma=1) = 0 \label{eq:R}.
\eea
In order to cast the system into the form discussed in Section \ref{sec:NumMeth}, we differentiate equations (\ref{eq:U}) and 
(\ref{eq:R}) with respect to $\tau$. The resulting equations are first order in $\tau$ and up to second order in $\sigma$, with third order mixed derivatives $U_{,\tau\sigma\sigma}$ appearing. The treatment of this system follows exactly along the lines described in Section \ref{sec:NumMeth}, with the following specifics:
\ben
	\item A first order reduction is not needed, i.e.~the SDIRK steps (\ref{eq:RKevol}) are performed with the implicit form (\ref{eq:K_implicit}). The concrete expressions can be found in appendix \ref{sec:SDIRK_frstTime}.
	\item As the system contains third order mixed derivatives $U_{,\tau\sigma\sigma}$, the last entry of 
	$F^A$ in (\ref{eq:K_implicit}) is not of algebraic form (as it would be for strictly first order systems in time and space)
	 but contains up to second order spatial derivatives. Therefore, in the limit $h\to 0$, eqns.~(\ref{eq:K_implicit})
	 become linear spatial PDEs of at most second order (for first order systems, it would be a linear {\em algebraic} system).
	 In the pseudo-spectral treatment of (\ref{eq:K_implicit}) for finite $h$, this fact is irrelevant.
\een
Eqs. (\ref{eq:U})-(\ref{eq:R}) possess the following time-independent equilibrium solution:
 \beq
 \rho_{\rm eq}(r) = \frac {\sin{(\pi r)}}{\pi r} , \quad U_{\rm eq}(r)= -2\left[ 1 +  \rho_{\rm eq}(r) \right],\quad w_{\rm eq} (r)=0. \label{eq:StarEquil}
  \eeq
We consider particular initial data which are related to the equilibrium solution. In particular, we choose $\rho(0,r)=\rho_{\rm eq}(r)$ and $R(0)=1$, which implies via (\ref{eq:U}) that we have to take $U(0,r)=U_{\rm eq}(r)$. We then perturb the star by specifying some profile $w(0,r)\ne 0$.

\subsubsection{Numerical results}
In~\cite{Hennig:2012zx}, much attention was paid to the results corresponding to a star with parameters $K=4.5\times 10^{-3}{\rm \frac{m^5}{kg s^2}}$ and $\rho_c=1.9891\times 10^{18}{\rm\frac{kg}{m^3}}$, which gives an equilibrium radius of $r_0=10.29~{\rm km}$. The perturbations were introduced by specifying a constant initial profile $w(0,r) = 400~{\rm s^{-1}}$. Moreover, each time domain was of the size $\Delta t = 8\times 10^{-5}{\rm s}$, and the spectral resolutions were chosen to be $N_{\sigma}=18$, $N_{\tau}=20$ (see fig.~1 in \cite{Hennig:2012zx}). In terms of the re-scaled dimensionless fields introduced in (\ref{eq:dimensionless}), these initial data correspond to a perturbation  $w(0,r)\approx 4.35\times 10^{-2}$ and the numerical solution was obtained in time domains of the size $\Delta t \approx 0.7$. 

Our first results focus on the question how stronger perturbations affect the performance of the Newton-Raphson method. After fixing the resolutions $N_{\sigma}=18$ and $N_{\tau}=20$ and parameterizing the perturbation in terms of $\epsilon$ by writing $w(0,r)=\epsilon$, we compare the number of iterations needed by the Newton-Raphson method as a function of $\epsilon$ for initial-guesses provided by: (i) the unperturbed solution (\ref{eq:StarEquil}), as described in~\cite{Hennig:2012zx} and (ii) the SDIRK method, as introduced in section~\ref{sec:SDIRK}. 

\begin{figure*}[h!]
\begin{center}
\includegraphics[width=8.4cm]{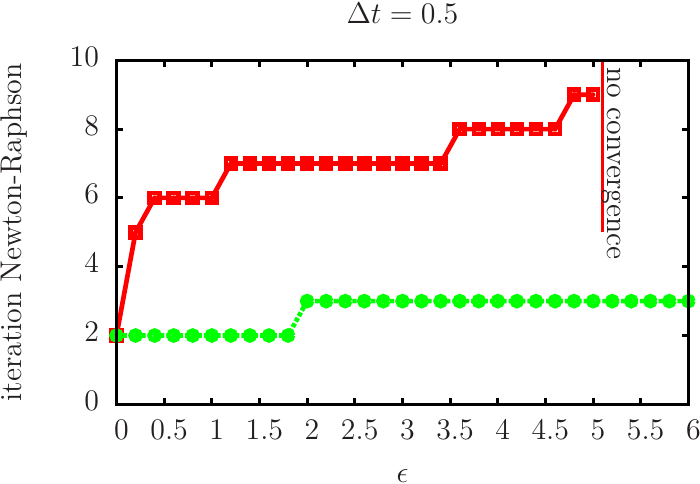}
\includegraphics[width=7.9cm]{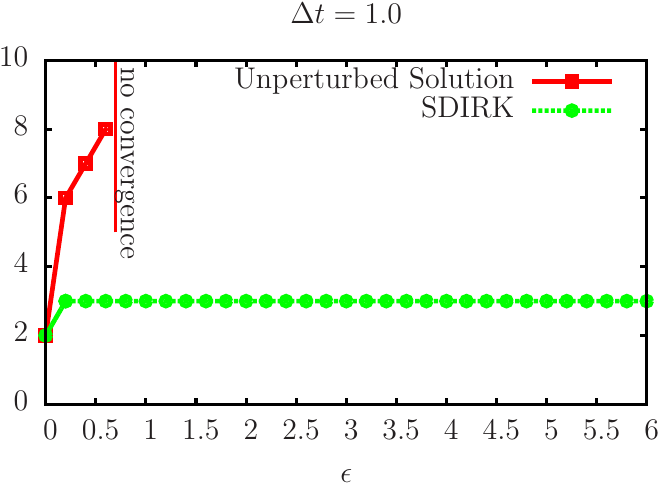}
\end{center}
\caption{Iterations needed by the Newton-Raphson scheme with two different initial guesses: (i) the unperturbed solution (red squares) and (ii) SDIRK method (green circles). While the first case requires many iterations ($\sim 6-9$), and might not converge at all, the second method provides convergence even for very strong perturbations and requires less iterations ($\sim 3$).}
\label{fig:NewtonConv}
\end{figure*}

Fig.~\ref{fig:NewtonConv} shows the results for two different time intervals ($\Delta t = 0.5$ in the left panel and $\Delta t = 1.0$ in the right panel). As might have been expected, for small time intervals, the unperturbed solution is a  good initial guess even for perturbations much stronger then $w \sim 10^{-2}$ (which was used in~\cite{Hennig:2012zx}). For $w \gtrsim 1$, the Newton-Raphson method requires  $6-9$ iterations to converge. It shows, however, no convergence for $w \gtrsim 5$. As we increase the time interval, the Newton-Rapshon fails already for $w \gtrsim 1$.

\begin{figure*}[h!]
\begin{center}
\includegraphics[width=8.0cm]{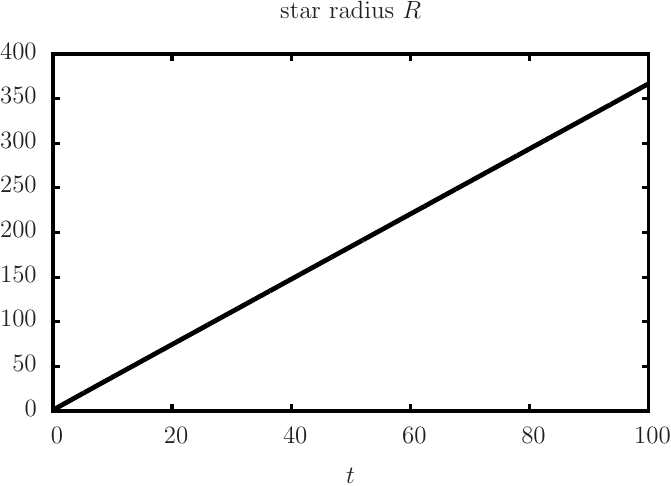}
\includegraphics[width=7.8cm]{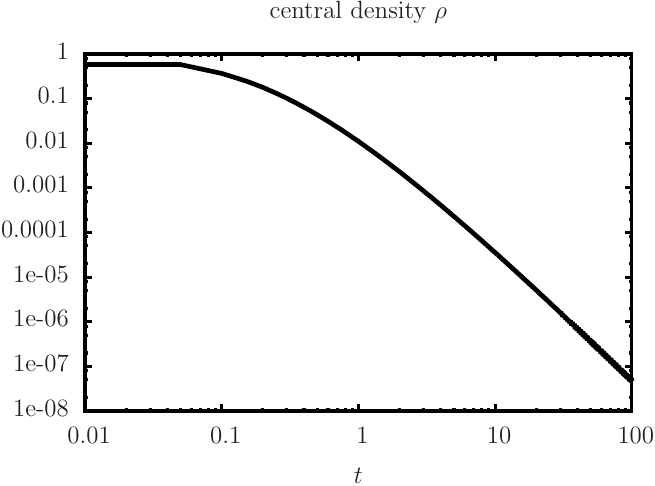}
\end{center}
\caption{Radius $R$ (left panel) and central density $\rho_{\rm c}=\rho|_{r=0}$ (right panel) of a star emerging from a large initial perturbation $w(0,r)\equiv \epsilon = 4$. Contrary to the weak perturbation regime, the star does not show oscillations but expands indefinitely. The radius $R$ grows larger with time, while the central density decreases steadily.}
\label{fig:StrongPert}
\end{figure*}

In this example it becomes apparent that, for a fully-spectral treatment of non-linear dynamical equations in terms of the Newton-Raphson method, a sufficiently good initial guess is essential. This statement applies to all time domains in which one wishes to solve the equations. In some cases, sufficiently good initial guesses are available through a known nearby solution (in the example for weak perturbations and small time domains). However, in general an efficient  construction of good initial guesses becomes an issue. We find that the SDIRK-method described in section~\ref{sec:SDIRK} is an excellent tool to solve this issue. For the computations corresponding to the green circles in Fig.~\ref{fig:NewtonConv}, the Newton-Rapshon was fed with the initial guess determined through the SDIRK-method. One can see that, no matter how strong the perturbation or how large the time interval is, the Newton-Rapshon method always converges quickly within just a few iterations ($\sim 3$) to the spectral solution. In Fig.~\ref{fig:StrongPert} we present an example of a strong perturbation ($\epsilon=4.0$), which leads to an indefinitely growing star with gradually decreasing density. The figure depicts the star's radius $R$ and central density $\rho_{\rm c}=\rho|_{r=0}$ against coordinate time $t$. 

Note that sufficiently small perturbations lead to mild oscillations with moderate gradients of the fields. Particular strong perturbations as the one discussed in Fig.~\ref{fig:StrongPert} lead to indefinitely growing stars, again with moderate gradients. However, strongly perturbed oscillating stars develop large spatial gradients in the vicinity of the coordinate $\sigma = 1$ which, in a pseudo-spectral treatment, need to be handled with care. These large spatial gradients occur for perturbations in the range $0.1 \lesssim \epsilon \lesssim 3.0$. It is not our objective to further explore the physical properties of such strongly perturbed oscillating stars, but we mention that a possible way to deal with this feature is to introduce new coordinates that populates the grid points around the problematic value $\sigma = 1$ (this technique is explained in the next application, see section \ref{sec:Scalar}). Furthermore, as discussed in~\cite{Rezzolla_book:2013}, an alternative for the treatment of spherical stars would be the formulation of the equations in terms of a Lagrangian description.

\begin{figure*}[h!]
\begin{center}
\includegraphics[width=10.0cm]{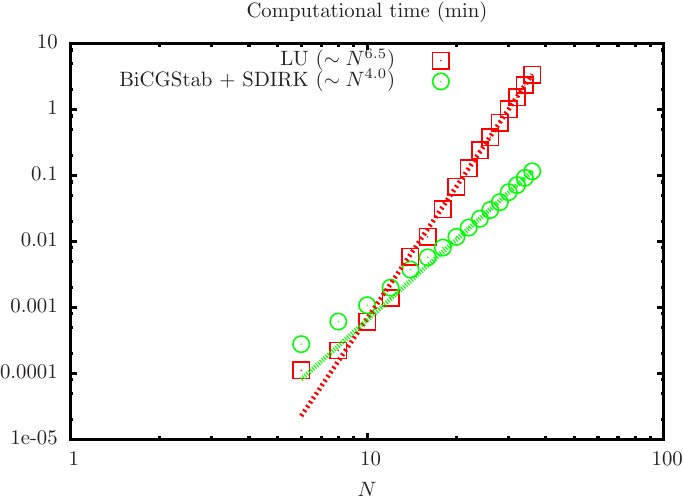}
\end{center}
\caption{Computation times for two different algorithms to solve the linear system (\ref{eq:NR_linSys2}) . Inverting the dense matrices with a LU-decomposition method (red squares) scales as $\sim N^6$. The iterative BiCGStab method equipped with an SDIRK pre-conditioner (green circles) is more efficient and scales as $\sim N^4$.}
\label{fig:CompTime}
\end{figure*}

We end this section with a discussion of the performance of the BiCGstab method equipped with an SDIRK pre-conditioner. In Fig.~\ref{fig:CompTime}, the computation times\footnote{The computation time depends on the particular machine on which the code is running. Here we are interested in its scaling with respect to the spectral resolution.}, needed for the solution of the linear system (\ref{eq:NR_linSys2}), are compared for two different algorithms, namely (i) the LU-decomposition and (ii) the BiCGstab-method with SDIRK pre-conditioner. We fix the size $\Delta t =1$ of the time domain and the perturbation parameter $\epsilon =1$ and prescribe mutual time and spatial resolutions, $N_\tau=N_\sigma = N$. As expected, the LU-decomposition scales as $\sim n_{\rm total}^3\sim N^6$. 
Within the BiCGstab-method, we solve (\ref{eq:K_implicit}) on each time slice $\tau_k$ ($k=0\ldots N_\tau$) by means of the LU-decomposition algorithm, thus obtaining a scaling of order $\sim n_{\sigma}^3 n_\tau\sim N^4$. \footnote{Note that we could do better if we were to apply an iterative scheme for the solution of (\ref{eq:K_implicit}), as explained at the end of section \ref{sec:SDIRK}. Indeed, iterative schemes become essential for more than one spatial dimension.}
While the difference in performance might still be acceptable for a (1+1)-problem, the solution of (\ref{eq:NR_linSys2}) in terms of the LU-decomposition becomes prohibitive when going to higher dimensions (see section \ref{sec:TeukEqn}). 

\subsection{The Teukolsky equation on hyperboloidal slices}\label{sec:TeukEqn}
We now abandon spherical symmetry and solve an axisymmetric dynamical problem in (2+1)-dimensions. More concretely, we consider the propagation of scalar, electromagnetic and gravitational perturbations on the background of a rotating black hole (the Kerr spacetime). This scenario is described by the Teukolsky equation~\cite{Teukolsky73}, which we treat here on hyperboloidal slices. Recently, studies of this kind were much-noticed, in particular when considering the behavior of the fields' decay at late times~\cite{Racz:2011qu,Harms:2013ib}.

%%%%%%%%%%%%%%%%%%%%%%%%%%%%%%%%%%%%%%%%%%%%%%%%%%%%%%%
\subsubsection{Hyperboloidal Foliation}\label{sec:HypFol}
We start with the presentation of hyperboloidal slices used in the remainder of this article. In general terms, we introduce a space-time coordinate system $\{ \tau, \sigma, \theta, \varphi\}$ with the following properties: 
\bit
	\item Through a conformal compactification, future null infinity $\scri$  is mapped to a finite coordinate radius $\sigma=\sigma_{\scri}$. 
	\item The surfaces $\tau =$constant are space-like and extend up to  future null infinity $\scri$.
\item The angular coordinates $\{ \theta, \varphi\}$ parametrize the closed 2-surfaces given by $\tau$=constant and $\sigma=$constant. 
\eit

A consequence of this choice is the fact that, in the  coordinates $\{ \tau, \sigma, \theta, \varphi\}$, the {\em physical} metric tensor $g_{\mu\nu}$ assumes a specific singular form, i.e.~it can be written as $g_{\mu\nu}=\Omega^{-2}\tilde{g}_{\mu\nu}$. Here, the {\em conformal} factor $\Omega$ is positive everywhere, apart from $\scri$, where it vanishes linearly, that is $\Omega|_\scri=0$ and $\partial_\sigma\Omega|_\scri\ne0$. The {\em conformal} metric $\tilde{g}_{\mu\nu}$ is regular everywhere up to and including $\scri$.

With focus on the numerical algorithm, we restrict the coordinate $\sigma$ to the domain $[\sigma_{\cal H}, \sigma_{\scri}]$, where $\sigma_{\cal H}$ describes the coordinate location of the event horizon. We thus obtain the horizon ${\cal H}$ and future null infinity $\scri$ as boundaries of our numerical grid. Through this choice the necessity of boundary conditions is removed, as $\sigma_{\cal H}$ and $\scri$ are null surfaces at which the characteristics of the Teukolsky equation point outside\footnote{In the case of dynamical space times, the inner boundary would correspond to the apparent horizon (see discussion in\cite{Schinkel:2013zm}). Here, the event and apparent horizon coincide.}.

The details for constructing hyperboloidal slices within the Kerr spacetime is presented in \cite{Schinkel:2013zm}. 
In short, one starts with the Kerr solution in horizon penetrating coordinates $\{v, r, \theta, \varphi \}$ (Kerr coordinates) and introduces $(\tau, \sigma)$ via
\bea
v &=& 4M \left[ \tau +  \left( \frac{1}{\sigma}  - \ln{\sigma} \right) \right]  \label{eq:HypCoord1} \\
r &=& \frac{2M}{\sigma}.  \label{eq:HypCoord2}
\eea
As a result, one gets the conformal line element $d\tilde{s}^2  =  \Omega^{2}ds^2$ with

\[\Omega = \frac{\sigma \sqrt{\eta}}{4M},\]

\bea
d\tilde{s}^2 & = &  - \sigma^2(\eta -\sigma)d\tau^2  + (\eta - 2\sigma^2\beta)d\tau d\sigma + \beta(1+\sigma)d\sigma^2   \nonumber \\
 &  - &  \sigma^3\sin^2{\theta}\frac{a}{2M} d\tau d\varphi + \frac{a}{2M}\frac{\sin^2{\theta}}{4}\left[\eta + 2\sigma(1+\sigma)\right] d\sigma d\varphi \nonumber \\ 
 &+&  \frac{\eta^2}{4}d\theta^2  + \frac{\sin^2{\theta}}{4}  \left\{ 1+ \frac{a^2\sigma^2}{4M^2}\left[ 1 + \eta + (\sigma -1)\sin^2{\theta}\right]   \right\} d\varphi^2  \label{eq:KerrHyp}
 \eea
 and \[ \eta = 1+\frac{a^2\sigma^2}{4M^2}\cos^2{\theta}, \quad \beta = 1 -  \frac{a^2\sigma}{4M^2}\cos^2{\theta} .\]
In the coordinates $\{ \tau, \sigma, \theta, \varphi\}$, future null infinity is given by $\sigma_{\scri} = 0$, whereas the horizon is located at $ \sigma_{\cal H} = 2/(1+\sqrt{1 - a^2/M^2})$.

A word should be said about the choice of our time coordinate $\tau$. Note that it scales as $\tau\sim v/(4M)$, while in~\cite{Racz:2011qu,Harms:2013ib} a time coordinate behaving as $T\sim v$ has been used. In the next sections we will show results running up to $\tau = 1000$. In terms of the coordinate $T$ utilized in~\cite{Racz:2011qu,Harms:2013ib}, it would correspond to data up to $T=4000M$, which is at least two times bigger than the time period of the long run simulations presented therein.
 
 \subsubsection{Teukolsky Equation}\label{sec:Teukolsky}
The equation describing the dynamics of a perturbation $X$ in a background given by the Kerr solution was derived by Teukolsky~\cite{Teukolsky73}. Originally written in Boyer-Lindquist coordinates $\{ t,r,\theta, \varphi\}$, the axisymmetric Teukolsky equation reads in our hyperboloidal coordinates $\{ \coord \}$ [see eqs.~(\ref{eq:HypCoord1}) and (\ref{eq:HypCoord2})]

\bea
\label{eq:IrregTeukolsky}
&-&  \left\{ 1+\sigma -\frac{a^2}{4M^2}\left[\sigma(1+\sigma) + \frac{\sin^2\theta}{4}\right]\right\} \frac{\partial^2 X}{\partial\tau^2}
+\left[ 1-2\sigma^2+\frac{a^2\sigma^2}{4M^2} \left(1+ 2\sigma \right) \right]\left(\frac{\partial X}{\partial\tau}\right)_{,\sigma} \nn \\
 &-& \left\{ \frac{1}{\sigma} -\frac{a^2\sigma^2}{4M^2} + \lambda \left[\frac{2}{\sigma} - \frac{2-\sigma}{\delta} \left( 1-\frac{a^2\sigma}{4M^2}\right)   -i \frac{a}{2M}\frac{\cos\theta}{\sigma \delta}\right]\right\}\frac{\partial X}{\partial\tau}  \\
  &+& \delta\sigma^2 X_{,\sigma\sigma} - \sigma\left[ \sigma\left(1-\frac{a^2\sigma}{2M^2} \right) + \lambda(2-\sigma)\right]X_{,\sigma}\nn
+ X_{,\theta\theta} +  X_{,\theta}\cot\theta  + \lambda\left(1-s \cot^2\theta \right)X = 0
\eea
with
\[\delta  = 1-\sigma + \frac{a^2\sigma^2}{4M^2}.\]
The parameter $s$ is related to the field's spin and it specifies the type of perturbation: scalar ($s=0$), electromagnetic ($s=\pm1$) or gravitational ($s=\pm2$). Note that eqn.~(\ref{eq:IrregTeukolsky}) degenerates for $\sigma = 0$ and $\delta =0$ (representing $\scri$ and  horizon $\sigma=\sigma_{\cal H}$,  respectively), as well as for $\sin\theta=0$. 

In order to compare our numerical results with those available in the literature, we follow the corresponding treatment in~\cite{Harms:2013ib} and consider the auxiliary field 

\[\psi(\coord) = \sigma^{-(1+2s)}\delta^{s}X(\coord)\] 
which was introduced to remove the degeneracies of the equation with respect to the $\sigma$-direction. As explained in Section \ref{sec:SpecCode}, we stick with the coordinate $\theta$ because fields with spin $s=\pm1$ possess terms proportional to $\sin\theta$ and are not regular in terms of the coordinate $\mu=\cos\theta$.

In terms of the new field $\psi$ the wave equation (\ref{eq:IrregTeukolsky}) reads

\bea
\label{eq:RegTeukolsky}
&-&  \sin^2\theta\left\{ 1+\sigma -\frac{a^2}{4M^2}\left[\sigma(1+\sigma) + \frac{\sin^2\theta}{4}\right]\right\}  \frac{\partial^2\psi}{\partial\tau^2}
+\sin^2\theta\left[ 1-2\sigma^2+\frac{a^2\sigma^2}{4M^2} \left(1+ 2\sigma \right) \right]\left(\frac{\partial\psi}{\partial\tau}\right)_{,\sigma} \nonumber \\
&-&\sin^2\theta\left[ 2\sigma -\frac{a^2\sigma}{4M^2} (1+3\sigma)- s\left(  1-\sigma - i \frac{a \cos\theta}{2M}  \right) \right] \frac{\partial\psi}{\partial\tau} \nonumber \\
 &+&\sin^2\theta \delta\sigma^2  \psi_{,\sigma\sigma} + \sin^2\theta\sigma\left[ 2-3\sigma + \frac{a^2\sigma^2}{M^2} + s\left( 2-\sigma \right)\right] \psi_{,\sigma} - \sin^2\theta\sigma \left( 1 + s -\frac{a^2\sigma}{2M}\right)\psi \nonumber \\
&+&\sin^2\theta \psi_{,\theta \theta} + \cos\theta\sin\theta \psi_{,\theta} + s\left(\sin^2\theta-s \cos^2\theta \right)\psi = 0.
\eea
Note that eqs.~(\ref{eq:IrregTeukolsky}) and (\ref{eq:RegTeukolsky}) are complex for $a\ne0$, so that the solution is complex as well. Numerically, we consider 
\beq 
\label{eq:psi_ReIm}
\psi(\coord) = \psi_{\rm R}(\coord) + i \psi_{I}(\coord)
\eeq and evolve the two coupled equations that results from substituting eq.~(\ref{eq:psi_ReIm}) into (\ref{eq:RegTeukolsky}).
Initial data for solutions of eqn.~(\ref{eq:RegTeukolsky}) are typically taken to be of the form 
\bea
\psi_{\rm initial}(\spacecoord) &=& F(\sigma)\,{}_s\!Y_{\ell' 0}(\theta) \nonumber \\
\left(\frac{\partial\psi}{\partial\tau}\right)_{\rm initial}(\spacecoord) &=& G(\sigma)\,{}_s\!Y_{\ell' 0}(\theta),
\eea 
where ${}_s\!Y_{\ell 0}(\theta)$ are the so-called spin weighted $s$ spherical harmonics~\cite{Goldberg:1967}. As a simple, though significant example, we will restrict ourselves to the radial profiles $F(\sigma)=1$ and $G(\sigma)=0$ (i.e., the initial data are purely real).

%%%%%%%%%%%%%%%%%%%%%%%%%%%%%%%%%%%%%%%%%%%%%%%%%%%%%%%%%
\subsubsection{The late time behavior: an introduction}\label{sec:Scalar}
When solving the Teukolsky equation on hyperboloidal slices, much attention has been paid to the late time behavior of the fields. We introduce into the subject and discuss some technical issues regarding the accuracy of the method by restricting ourselves to the most simple version of the Teukolsky equation (\ref{eq:RegTeukolsky}) obtained for the parameters $a=0$ and $s=0$. The resulting equation,
\bea
\label{eq:ScalarSchwarz}
&-&  \sin^2\theta\left( 1+\sigma \right)  \frac{\partial^2\psi}{\partial\tau^2}+\sin^2\theta\left( 1-2\sigma^2\right)\left(\frac{\partial\psi}{\partial\tau}\right)_{,\sigma} 
-2\sigma\sin^2\theta\frac{\partial\psi}{\partial\tau} \nonumber \\
 &+&\sin^2\theta (1-\sigma)\sigma^2  \psi_{,\sigma\sigma} + \sin^2\theta\sigma\left( 2-3\sigma \right) \psi_{,\sigma} - \sin^2\theta\sigma\psi \nonumber \\
&+&\sin^2\theta \psi_{,\theta \theta} + \cos\theta\sin\theta \psi_{,\theta} = 0,
\eea
describes the propagation of a scalar field on the background given by a non-rotating black hole (the Schwarzschild space-time).
From eqn.~(\ref{eq:ScalarSchwarz}) it follows that $\psi$ depends, for the case $a=s=0$, effectively only on $\mu=\cos\theta$. Although we do not take this property into account within our  numerical solution procedure (as explained above), we need to ensure that the initial data $\psi_{\rm initial}$ and $(\partial\psi/\partial\tau)_{\rm initial}$ possess this property. 

\begin{figure*}[h!]
\begin{center}
\includegraphics[width=10.0cm]{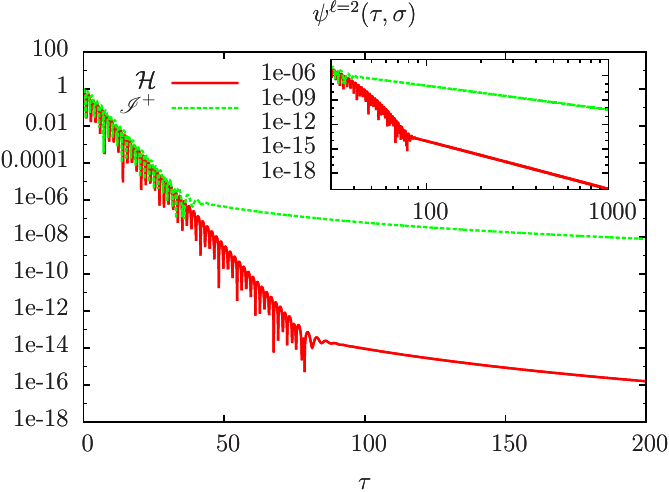}
\end{center}
\caption{Evolution of the scalar field on the Schwarzschild background given by eq.~(\ref{eq:ScalarSchwarz}). The field value at the horizon ${\cal H}$ and at future null infinity are shown, respectively, by the red continuos line and the green dotted plot. The figure displays the typical initial ring-down phase, with decay and oscillating time scales given by the quasi-normal modes, followed by an inverse power law decay $\tau^{-\alpha}$.}
\label{fig:ScalarFieldSchw}
\end{figure*}

For the initial data  $\psi_{\rm initial}(\spacecoord) = P_2(\cos\theta)$ and $(\partial\psi/\partial\tau)_{\rm initial}(\spacecoord) = 0$ (where $P_\ell(\cos\theta)$ are Legendre polynomials), fig.~(\ref{fig:ScalarFieldSchw}) shows the time evolution of the ($\ell$=2)-mode $\psi^{\ell=2}(\tau, \sigma)$ (obtained after a projection of $\psi$  onto $P_{\ell=2}(\cos(\theta))$\footnote{The evolution was performed in time intervals of size $\Delta \tau = 2$ with the spectral resolutions $N_\tau=N_\theta=45$ and $N_\sigma=55$ [see figs.~(\ref{fig:ScalarFieldSchw_ChebCoef}) and (\ref{fig:ScalarFieldSchw_ChebCoef2})].}.
Here one sees clearly the typical behavior of fields propagating on a black-hole space-time, namely, an initial ring-down phase, with oscillations and decay time scales given by the quasi-normal modes (QNM)~\cite{lrr-1999-2}, and a final tail, characterized by an inverse power law decay $\tau^{-\alpha}$~\cite{Price72}. 

\begin{figure*}[h!]
\begin{center}
\includegraphics[width=8.0cm]{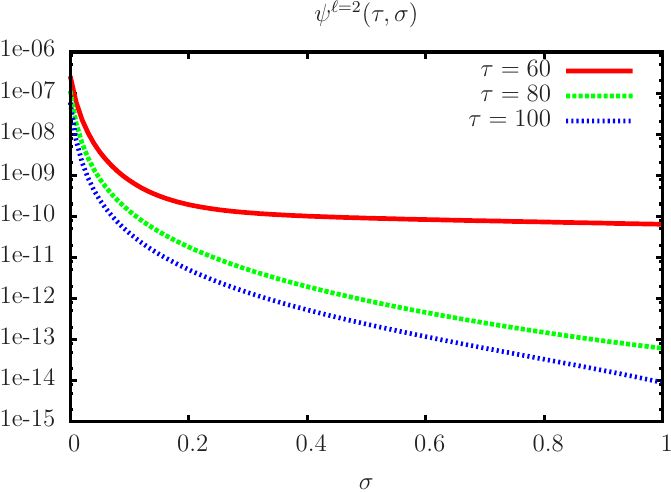}
\includegraphics[width=8.0cm]{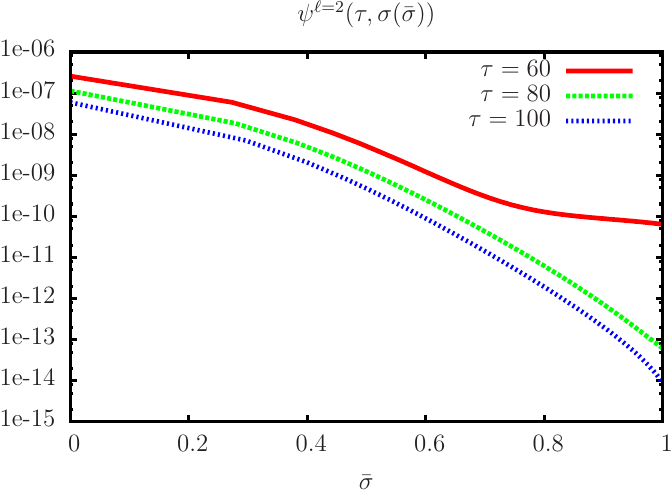}
\end{center}
\caption{Due to the differences in the field's decay rates at ${\cal H}$ and at future null infinity, the function $\psi(\coord)$ develops  a strong gradient with respect to the $\sigma-$direction. The left panel shows the field's behavior with respect to $\sigma$. Note that within the interval $[0,0.2]$ the field's value drops significantly (about 3-5 orders of magnitude). The right panel, on the other hand, presents the dependence with respect to the new coordinate $\bar{\sigma}$. The strong gradients are eliminated and we obtain a better spectral convergence.
}
\label{fig:ScalarFieldSchw_GradSigma}
\end{figure*}

\begin{figure*}[h!]
\begin{center}
\includegraphics[width=5.4cm]{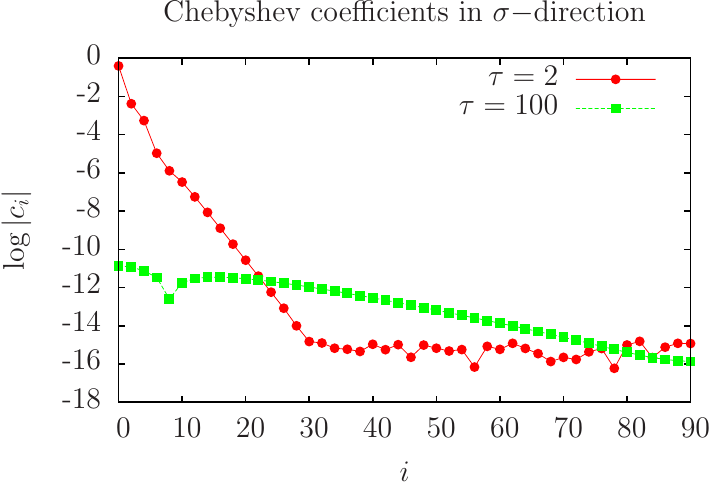}
\includegraphics[width=5.4cm]{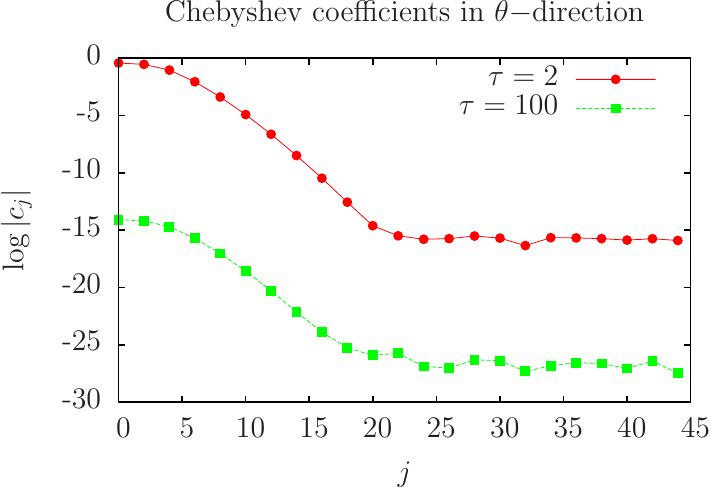}
\includegraphics[width=5.4cm]{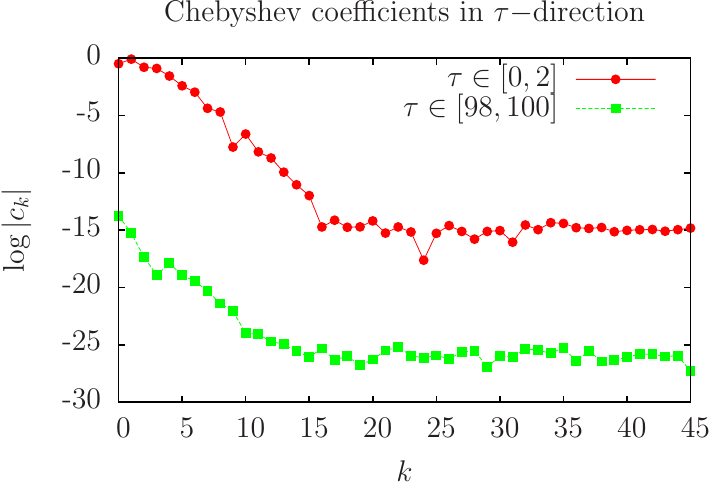}
\end{center}
\caption{Chebyshev coefficients of the function $\psi(\coord)$ with respect to $\sigma$, $\theta$ and $\tau$, respectively. The coefficients are calculated in the left panel for:  $\sigma \in [0,1], \theta = 0, \tau = 2$ (red circles) and $\tau = 100$ (green squares),  in the middle panel for: $\sigma =1,\,\theta \in [0,\pi], \tau = 2$ (red circles) and $\tau = 100$ (green squares ), and in the right panel for:  $\sigma=1, \theta = 0, \tau\in[0,2]$ (red circles) and $\tau\in[98,100]$ (green squares). The high resolution is needed only regarding the $\sigma$-direction.
 }
\label{fig:ScalarFieldSchw_ChebCoef}
\end{figure*}

\begin{figure*}[h!]
\begin{center}
\includegraphics[width=9.0cm]{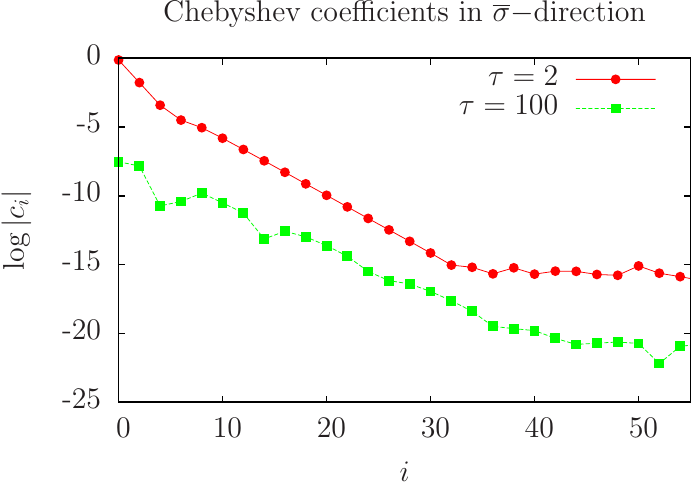}
\end{center}
\caption{Chebyshev coefficients of the function $\psi(\tau,\sigma(\bar\sigma),\theta)$ with respect to $\bar\sigma$ calculated for $\theta = 0, \tau = 2$ (red circles) and $\tau = 100$ (green squares) (as in Fig. \ref{fig:ScalarFieldSchw_ChebCoef}). The mapping $\sigma(\bar{\sigma})$ allows one to achieve the saturation with much fewer coefficients (to be compared with the left panel of fig.~\ref{fig:ScalarFieldSchw_ChebCoef}). At late times $\tau\sim 100$, the parameter in (\ref{eq:sigma_bar}) amounts via (\ref{eq:epsilon})  to $\kappa \sim 5$.}
\label{fig:ScalarFieldSchw_ChebCoef2}
\end{figure*}

Note that the field decays much faster at the horizon than at $\scri$ (see  inset on the left panel of fig.~\ref{fig:ScalarFieldSchw}). This fact means that, with respect to the $\sigma$-direction, the function develops a strong gradient in the vicinity of $\scri$ (as shown in the left panel of fig.~\ref{fig:ScalarFieldSchw_GradSigma}). As a consequences of such strong gradients, the decay rate of the Chebyshev coefficents drops significantly, as can be seen in fig.~(\ref{fig:ScalarFieldSchw_ChebCoef}). Though they show the typical exponential decay, we note that, for late times, one needs a rather high resolution with respect to the $\sigma$-direction, while the grid points in the $\tau$ and $\theta$-directions could be fixed to be around $\sim20$.

A way to treat functions with steep gradients in spectral methods is described in \cite{Meinel:2008}. The function 
\beq\label{eq:f_aux_func} f(\sigma) = \frac{\epsilon}{\epsilon + \sigma}\eeq models the behaviour presented in Fig.~(\ref{fig:ScalarFieldSchw_GradSigma})  when $\epsilon \ll 1$. In particular, its $m$th derivative at $\sigma =0$ goes as $\epsilon^{-m}$ and becomes very large in the $\epsilon \ll 1$ regime. With the introduction of a new coordinate $\bar{\sigma}$ through the relation 
\beq\label{eq:sigma_bar} \sigma = \sigma_{\cal H}\frac{\sinh{\kappa \bar{\sigma}}}{\sinh{\kappa}}, \quad {\rm with} \quad \kappa = |\ln\epsilon|, \eeq 
the $m$th derivative of the function $\bar{f}(\bar{\sigma})= f(\sigma(\bar{\sigma}))$ scales merely as $(\ln\epsilon)^{m}$, and we achieve machine precision with a moderate number of grid points.

Here, we take advantage of this analytic mesh-refinement. In particular, we set in our code the order of magnitude of $\epsilon$ at the end of each time domain to be 
\beq\label{eq:epsilon}\epsilon = \min\{1,\bar{\epsilon}\},\quad\mbox{with}\,\, \bar\epsilon = \min_{ \theta} \frac{ \psi_{|\scri}}{ \psi_{,\sigma}{}_{|\scri}}.\eeq 
Once $\epsilon$ is fixed, we also increase the resolution $N_\sigma$ utilizing the properties of the auxiliary function (\ref{eq:f_aux_func}). More concretely, we look at the Chebyshev coefficients $c^{(f)}_k$ of $f(\sigma)$ and choose $N_\sigma$ such that $|c^{(f)}_{N_\sigma}| \approx 10^{-15}$. In Fig.~\ref{fig:ScalarFieldSchw_ChebCoef2}, the resulting Chebyshev coefficients (emerging for $\kappa\sim 5$) of our scalar field in the Schwarzschild background with respect to the new coordinate $\bar\sigma$ are presented. As expected, the saturation is achieved with a substantially lower resolution.

Note that the new mapping $\sigma=\sigma(\bar{\sigma})$ populates the grid points around the region with the steep gradient, resulting in much smaller step sizes in $\sigma$-direction. This occurs for late times when high resolution in the $\tau$-direction is not needed. Hence, this mesh-refinement would face severe drawbacks when applied within a code utilizing an explicit time integrator, since the high resolution in the $\sigma$-direction would require an even higher time resolution. Our fully spectral method, though, can easily handle the difference regarding the two resolution scales and imposes no restriction whatsoever on the relation between $N_{\tau}$ and $N_{\sigma}$.   

%%%%%%%%%%%%%%%%%%%%%%%%%%%%%%%%%%%%%%%%%%%%%%%%%%%%%%%%%
\subsubsection{The general case: time evolution and Chebyshev coefficients}\label{sec:Gen_Case}

\begin{figure*}[h!]
\begin{center}
\includegraphics[width=8.1cm]{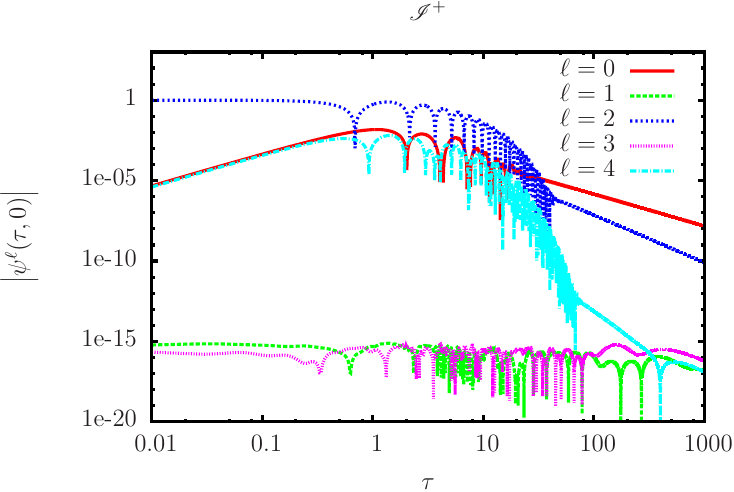}
\includegraphics[width=8.1cm]{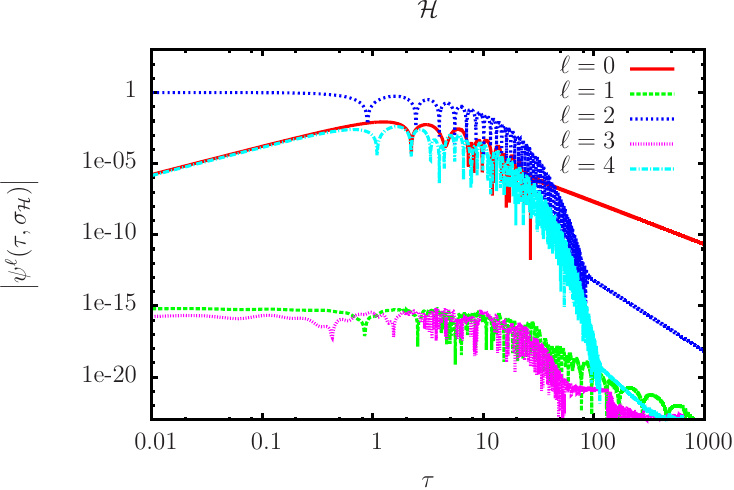}
\end{center}
\caption{Time evolution of the field's $\ell$-mode projection on the spin-weighted spherical harmonics ${}_0\!Y_{\ell 0}(\theta)$. The field's spin parameter is $s=0$, and the initial data are characterized by a purely angular ($\ell'$=2)-mode. The left and right panel depict, respectively, the field at future null infinity and ${\cal H}$. Note that only modes with the same parity as the initial one are excited. In particular, in the course of the entire evolution, the strengths of the modes $\ell=1$ and $\ell=3$ remain of the order of machine precision.}
\label{fig:s0_l2}
\end{figure*}

In the general case described by eq.~(\ref{eq:RegTeukolsky}), the field's time evolution possesses very similar properties as the one in the Schwarzschild case discussed above: it shows an initial quasi-normal ring-down phase,  followed by a late time tail decay. However, contrary to case $a=0$, the spin weighted spherical harmonic ${}_s\!Y_{\ell 0}(\theta)$ do not constitute a natural basis to the solutions of  eq.~(\ref{eq:RegTeukolsky}). Consequently,   initial data prescribed with a single $\ell'$-mode will, in the course of time, excite other modes with $\ell \ne \ell'$.  Fig. (\ref{fig:s0_l2}) shows a typical solution for the field which was written as\footnote{Figures (\ref{fig:s0_l2}) and (\ref{fig:s1l2l}) show the norm $\left|\psi^{\ell}(\tau, \sigma)\right| = \sqrt{ \left|\psi^{\ell}_{\rm R}(\tau, \sigma)\right|^2 + \left|\psi^{\ell}_{\rm I}(\tau, \sigma)\right|^2  }$.}
\[\displaystyle \psi(\coord) = \sum_{\ell} \psi^{\ell}(\tau, \sigma) {}_sY_{\ell 0}(\theta).\]
 This projection on the spin weighted spherical harmonics is efficiently and accurately performed by means of the techniques described in~\cite{Huffenberger:2010hh}.  Here, the specific angular momentum parameter was chosen to be $a/M=0.9$, and the initial data are characterized by $\ell' =2$ and $s=0$. 

For vanishing spin parameter $s=0$, the symmetry of the Teukolsky equation implies that the excited $\ell$-modes have the same parity as the initial $\ell'$-mode. Consequently, in the example considered, the odd modes $\ell=1,3,\ldots$ do not occur. Fig.~\ref{fig:s0_l2} shows that during the entire numerical evolution the strengths of the modes $\ell=1$ and $\ell=3$ remain of the order of machine precision.

\begin{figure*}[h!]
\begin{center}
\includegraphics[width=5.4cm]{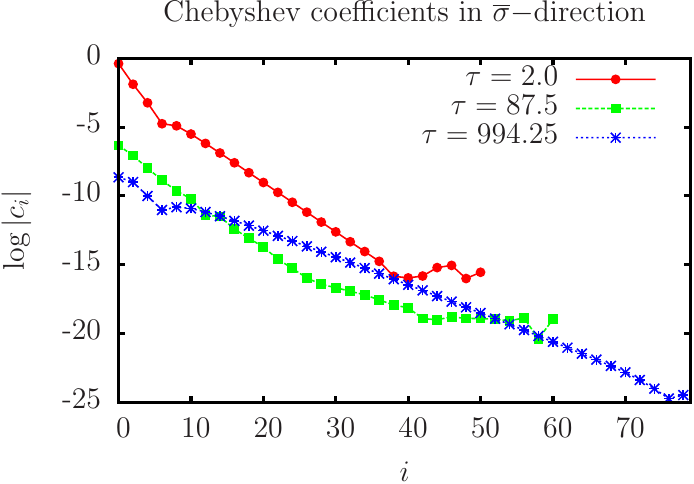}
\includegraphics[width=5.4cm]{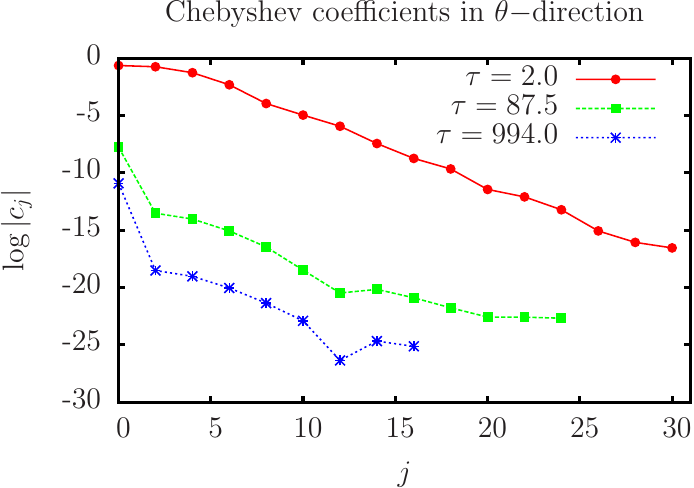}
\includegraphics[width=5.4cm]{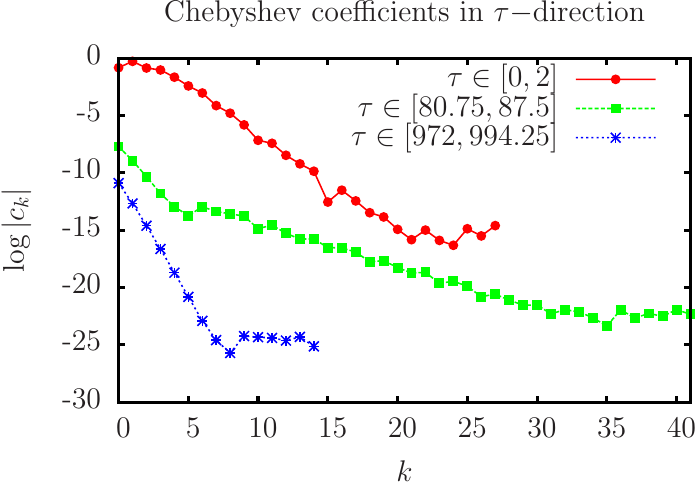}
\end{center}
\caption{For $\theta=0$, $\tau = 2$ (red circles), $\tau = 87.5$ (green squares) and $\tau = 994.25$ (blue stars), the left panel shows Chebyshev coefficients with respect to $\bar\sigma$. The middle panel presents coefficients with respect to the $\theta$-direction (here taken at the horizon $\sigma=\sigma_{\cal H}$). The right panel depicts coefficients with respect to the $\tau$-direction (again taken at the horizon $\sigma=\sigma_{\cal H}$ and $\theta=0$) for the intervals $\tau\in[0,2]$ (red circles), $\tau\in[80.75, 87.5]$ (green squares) and $\tau\in[972, 994.25]$ (blue stars). At late times, high resolution is needed with respect to the $\bar\sigma$-coordinate while the $\tau$-direction can be resolved with a rather small number of points. The large difference in the two resolutions neither affects the stability of the code, nor the high precision achieved by the spectral method.
}
\label{fig:ChebExampleTeukSol}
\end{figure*}

We focus now on the accuracy of our numerical results. Fig.~\ref{fig:ChebExampleTeukSol} depicts Chebyshev coefficients corresponding to the solution displayed in fig.~\ref{fig:s0_l2}. We consider separately the three different phases of the evolution:
\ben
\item During the initial quasi-normal ring-down phase, the equation is solved within rather small time intervals. Here we take, for instance, $\tau\in[0,2]$. Saturation of the numerical solution is reached for a resolution of the order $N_\sigma = 50$, $N_\theta=31$ and $N_\tau = 27$. As steep gradients do not yet develop, we may choose $\sigma=\sigma_{\cal H}\bar\sigma$ (i.e.~$ \kappa=0$). 
\item Around a coordinate time of $\tau \sim 90$, some modes enter the tail decay phase, while others are still ringing down. This behavior  can still be resolved with a resolution of  $N_\sigma \sim 60$. For the transformation parameter in (\ref{eq:sigma_bar}), a value $\kappa \sim 5$ emerges via (\ref{eq:epsilon}).  For the $\theta$-coordinate a smaller resolution $N_\theta\sim 25$ suffices. An accurate resolution with respect to the time direction requires values of $N_\tau\sim 40$. It is, however, important to note that the time interval is now larger than the one for the initial quasi-normal ring-down phase. Specifically, the solution is obtained for $\tau \in [80.75, 87.5]$.
\item At late times, the time interval can be chosen to be even larger ($\tau \in [972, 994.25]$). Only a small number of grid points with respect to the $\tau$-direction are required to reach saturation ($N_{\tau}\sim 15$). As the steep spatial gradients are extremely accentuated in this late phase, the transformation parameter in (\ref{eq:sigma_bar}) amounts via (\ref{eq:epsilon}) to $\kappa \sim 7$,
and the resolution with respect to the $\sigma$-direction needs to be chosen moderately large, $N_\sigma \sim 80$.  In an explicit time integration scheme, the very small spatial step sizes in the vicinity of $\scri$ would impose even smaller time steps, through which the method would become extremely expensive. In fact,~\cite{Harms:2013ib} reports obstacles for obtaining stable evolutions with a spectral code in the spatial direction when combined with an explicit 4th-order Runge-Kutta time integrator. 
\een

%%%%%%%%%%%%%%%%%%%%%%%%%%%%%%%%%%%%%%%%%%%%%%%%%%%%%%%%%
\subsubsection{Power law index}\label{sec:Pow_index}

Recently, there has been some effort to classify the so-called power index $\alpha$ of the late time tail decay of the excited $\ell$ mode, $\psi^\ell\sim \tau^{-\alpha}$, in dependence of the type of prescribed initial data. Many results in the context of hyperboloidal foliations were presented in \cite{Racz:2011qu} for the scalar field ($s=0$) and in \cite{Harms:2013ib} for electromagnetic ($s=\pm 1$) and gravitational ($s=\pm 2$) perturbations. Here we look into this matter in order to obtain a severe test for the robustness of our method and, furthermore, to provide an important cross-check for the validation of previous results.

From the results presented in section \ref{sec:Gen_Case} one can read off the power index $\alpha$ of the tail decay. The values shown in Table \ref{tab:powerindex} coincide with the ones given in \cite{Racz:2011qu}. Moreover, we succeeded in the determination of the index corresponding to the field's decay at the horizon ${\cal H}$ for the excited ($\ell$=2)-mode, which is absent in \cite{Racz:2011qu}. However, the ($\ell$=4)-mode could not be resolved, as the tail decay sets in only after the mode's strength has dropped below the round-off error, see fig.~\ref{fig:s0_l2}.

\begin{figure*}[h!]
\begin{center}
\includegraphics[width=5.4cm]{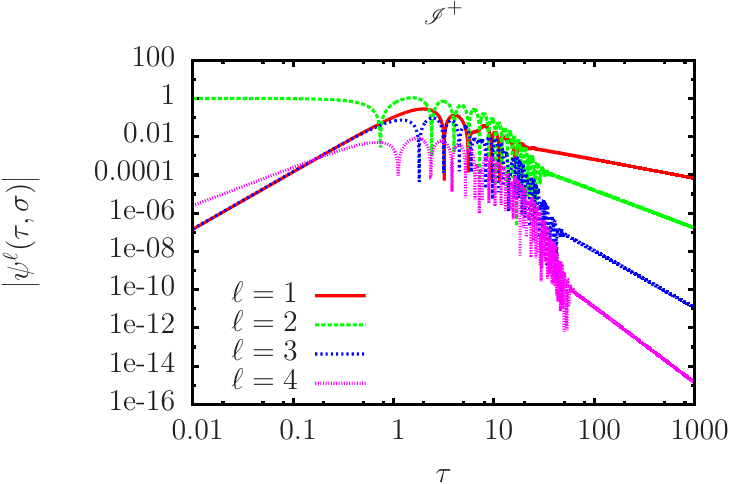}
\includegraphics[width=5.4cm]{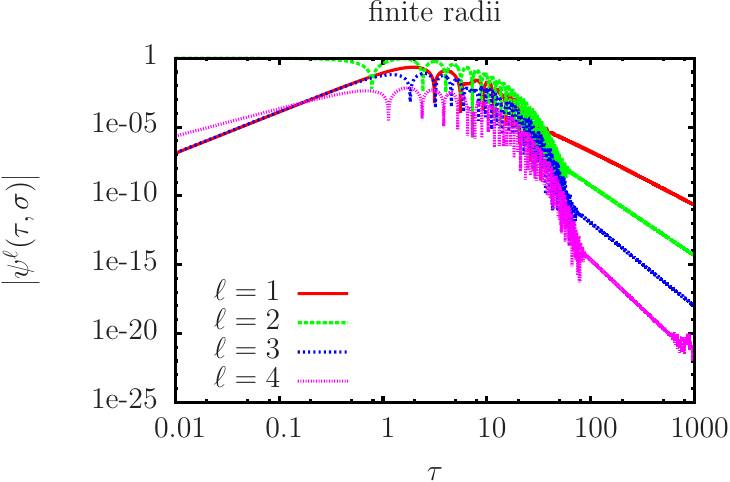}
\includegraphics[width=5.4cm]{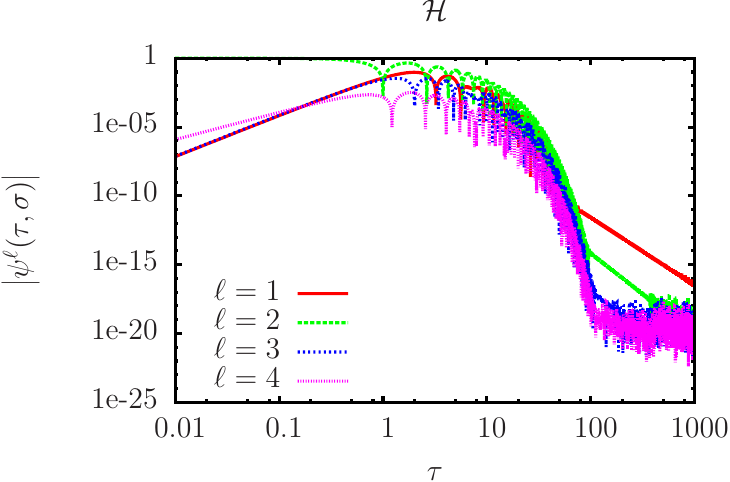}
\end{center}
\caption{Time evolution of the field's $\ell$-mode projection onto the spin-weighted spherical harmonics ${}_1\!Y_{\ell 0}(\theta)$.  The field has a spin parameter $s=+1$, and the initial data are characterized by a purely angular ($\ell'$=2)ömode.  Left, middle and right panel depict, respectively, the field at future null infinity, at finite radii (here, in particular, at $\sigma=0.5$) and at ${\cal H}$. Specific values are shown in Table \ref{tab:powerindex}.
}
\label{fig:s1l2l}
\end{figure*}

Filling up the table for other $\ell'$ modes and for fields with spin weight $s\neq 0$ would be a rather extensive exercise that would go beyond our objective. Nonetheless we mention that we performed several simulations for different configurations and the results agree with the ones presented in \cite{Racz:2011qu, Harms:2013ib}. In fig.~\ref{fig:s1l2l} we show especially the results for spin parameter $s=+1$. 

The reason for choosing a situation with positive spin parameter $s$ emerges from an interesting result in~\cite{Harms:2013ib}. For fields with $s\leq0$, the power index can only assume two different values: one, if the field is considered at $\scri$ and another one for measurements elsewhere, {\em including} the horizon ${\cal H}$. If, on the contrary, the spin parameter $s$ is positive, then the field decays
at the horizon ${\cal H}$ with a different index as compared to the decay in the bulk. That is to say that the field decays with three different rates. The results obtained for a representative situation  are given in Table \ref{tab:powerindex}. Note that we observe again that the power index $\alpha$ corresponding to some modes could not be resolved. A sophisticated determination of $\alpha$ in these cases requires higher internal precision or a specifically designed method.

\begin{table}[h]
\begin{center}
\caption{Power index $\alpha$ of the tail decay $\psi^\ell\sim\tau^{-\alpha}$ for spin weighted fields $s=0$ and $s=+1$, with an initial ($\ell'$=2)-mode and radial profile  $F(\sigma)=1$ and $G(\sigma)=0$. For $s=0$, the first entry corresponds to the decay rate measured at finite radii (including ${\cal H}$) and the second one to decays at future null infinity. For $s=+1$ three different decay rates (measured at ${\cal H}$, finite radii outside ${\cal H}$, and future null infinity) are shown. A bar "--" means that the corresponding mode does not occur, while a cross "$\times$" denotes failure in the determination of $\alpha$ due to restricted internal precision.\vspace*{2mm}}
\label{tab:powerindex}
\begin{tabular}{|c|c||c|c|c|c|c|}
\hline
{$s$}&\mbox{$\ell'$}&$\ell=0$ & $\ell=1$ & $\ell=2$& $\ell=3$ & $\ell=4$ \\ 
\hline
0 & 2&  $3\quad2$ & -- & $5\quad 3$ & -- & $\times\quad 5$ \\ 
\hline
+1 & 2&   --  & 5\quad 4 \quad 1  & $ 6\quad 5 \quad 2$ & $7\quad 6\quad 3$ & $\times\quad 7 \quad 4$ \\ 
\hline
\end{tabular} 
\end{center}
\end{table}

%%%%%%%%%%%%%%%%%%%%%%%%%%%%%%%%%%%%%%%%%%%%%%%%%%%%%%%%%
\subsubsection{Schwarzschild case: power index splitting for positive spin parameter}
\begin{figure}[h!]
\begin{center}
\includegraphics[width=10.0cm]{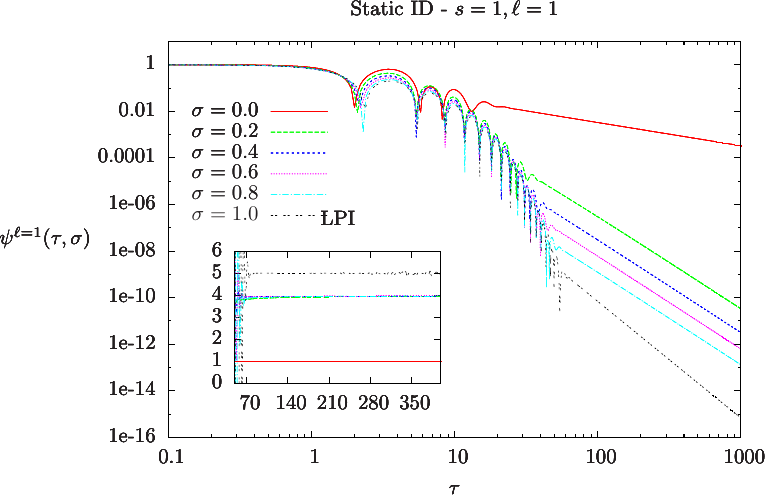}
\end{center}
\caption{Evolution of the ($\ell$=1)-mode of a spin ($s$=+1)-field on a Schwarzschild background ($a=0$). As in situations in the Kerr background, for $s>0$ the tail decay splits into three different values which depend on the location where the field is measured. At future null infinity we see a decay rate $\psi^{\ell=1}\sim\tau^{-1}$, at ${\cal H}$ it scales as $\psi^{\ell=1}\sim\tau^{-5}$ and at finite radii outside ${\cal H}$ we have $\psi^{\ell=1}\sim\tau^{-4}$. The values can be read off from the inset, which shows the {\em local power index} "LPI", defined by $p(\tau)=-\log\psi^\ell/\log\tau$ (tending asymptotically to the power index $\alpha$). The notation "Static ID" refers to initial data with $(\partial\psi/\partial\tau)_{\rm initial}(\spacecoord) = 0$.}
\label{fig:TeukSolSchwarzSpinPlus}
\end{figure}

The occurrence of three different power indexes (as discussed in Section \ref{sec:Pow_index}) has been investigated in the literature for fields propagating in a Kerr background. As a side result of our studies we find that three different decay rates are also present in a Schwarzschild background ($a=0$). Fig.~\ref{fig:TeukSolSchwarzSpinPlus} shows the evolution of a field with spin parameter $s=+1$ where the initial data are given by $\psi_{\rm initial}(\spacecoord) = {}_1\!Y_{1 0}(\theta)$ and  $(\partial\psi/\partial\tau)_{\rm initial}(\spacecoord) = 0$. The inset displays the time evolution of the so-called {\em local power index}, 
$p(\tau)=-\log\psi^\ell/\log\tau$ which tends to the power index $\alpha$ as $\tau\to\infty$. For this particular example, we get for $\ell=1$ decay rates $\alpha = 1$ if the field is evaluated at $\scri$, $\alpha = 4$ if it is measured at finite radii outside ${\cal H}$, and $\alpha=5$ if it is considered at the horizon. Similar results are also obtained when the spin parameter amounts to $s=+2$. Due to the restricted internal precision, we observe again failures in the determination of $\alpha$ for specific modes.

%%%%%%%%%%%%%%%%%%%%%%%%%%%%%%%%%%%%%%%%%%%%%%%%%%%%%%%%%
\section{Conclusion}\label{sec:Conc}
We have presented a novel numerical technique for the solution of axisymmetric hyperbolic equations.  The scheme can handle regular as well as irregular equations and permits the treatment of free boundaries, which are unknown at the outset but need to be determined in the course of the evolution. The underlying algorithm is a {\em fully} pseudo-spectral method, i.e.~spectral decompositions of the field variables are performed with respect to both spatial {\em and} time coordinates. We retain the most prominent properties of spectral methods, that is, exponential convergence rate of the numerical solutions and saturation close to machine precision. We use the Chebyshev polynomials as basis functions for the spectral representations with respect to all directions and work with collocation points based on the Lobatto grid for the spatial directions and on the Radau grid for the time direction. 

The scheme is an extension of the work presented in~\cite{Hennig:2008af,Ansorg:2011xg, Hennig:2012zx} and it solves two issues which restricted the applicability of previous versions, namely, (i) the inversion of large dense matrices and (ii) the acquisition of a sufficiently good initial-guess for the Newton-Raphson method. In our scheme, an efficient  inversion of large dense matrices is performed using the iterative BiCGstab method which is equipped with a pre-conditioner based on a Singly Diagonally Implicitly Runge-Kutta (SDIRK) method. Moreover, the SDIRK-method solves in an efficient and stable manner the second issue, as it provides us with a good initial-guess.

We demonstrated the improvement in performance, as compared to the method presented in~\cite{Hennig:2012zx}, by solving again the equations describing a spherically-symmetric dynamical star in the Newtonian theory of gravity. With an initial guess constructed through the SDIRK-method, we found rapid convergence of the Newton-Raphson method, independent of the size of the perturbation away from the explicitly known static solution. Most importantly, the iterative BiCGStab method performs an efficient inversion of the dense matrix, which permits the treatment of dynamical problems with more than one spatial dimension. 

The algorithm is meant to solve, in the near future, the axisymmetric Einstein equations on hyperboloidal slices in a free evolution scheme. For this purpose, we have applied the method to compute a scenario describing perturbations around a rotating black hole. More concretely, we solved axisymmetric wave equations in (2+1)-dimensions on a space-time background given by the Kerr solution. Coordinates were chosen such that surfaces of constant time extend up to future null infinity. In addition, the space-time was conformally compactified so as to include future null infinity as a finite boundary of the numerical grid ("hyperboloidal slicing"). Moreover, the coordinates were taken to be regular near the black hole horizon ("horizon-penetrating"), and the horizon was used as another boundary of the numerical domain. In other words, the wave equations were considered inside a domain with boundaries at which the corresponding characteristics point outward. Consequently, no boundary conditions were required to be imposed.

We conclude with a word about the fact that our algorithm  can cope with very different resolutions regarding time and spatial coordinates. As described in \cite{Hennig:2008af,Ansorg:2011xg, Hennig:2012zx}, this feature is due to the implicitness of the scheme, realized in both the fully pseudo-spectral and the SDIRK-method. In particular, we were able to take relatively large time steps ($\Delta\tau \sim 100$) with a small number of grid points ($n_\tau\sim 10$), and, at the same time, a high resolution in the radial direction ($n_\sigma \sim 100$ for an interval $\Delta\sigma \sim 1$). The advantage of such an asymmetric grid scaling becomes apparent when solving the Teukolsky equation on the Kerr background. For fixed radius, the late time behavior of the wave field is characterized by a power law decay, which possesses a rapidly converging spectral representation. At fixed late time, however, the field is characterized by a strong gradient with respect to the radial direction, and a highly accurate representation requires therefore a dense mesh. The different resolution scales arising in this situation did not impose any restriction to the applicability of the scheme.  Rather, our numerical solutions of the Teukolsky equation reproduced correctly results which are published elsewhere. In particular, the splitting of the so-called local power index into three different values, which was observed in~\cite{Harms:2013ib}, was successfully reproduced and extended to the Schwarzschild space-time.

\section*{Acknowledgements}
It is a pleasure to thank Christian Lubich for many fruitful suggestions and Andreas Weyhausen for discussions and help when debugging the code. This work was supported by the DFG-grant SFB/Transregio 7 ``Gravitational Wave Astronomy''.

\section{Appendix}\label{App}
\subsection{First order in time SDIRK method}\label{sec:SDIRK_frstTime}
It is straightforward to adapt the Runge Kutta method from section~\ref{sec:SDIRK} to system of PDEs, which are first order in time. Similar to eq. (\ref{eq:RKgenForm}), we consider the system in the form
\beq
\label{eq:RKgenForm2}
F^A\left(\tau, \left\{ X^B\right\}, \left\{ \frac{\partial X^B}{\partial\tau} \right\} \right) = 0.
\eeq

Then, according to eq.~(\ref{eq:RKevol}), the time evolution of $X^A$ is given after $K^A_{(i)}$ is obtained via 
\[
F^A\left(\tau_k + hc_{(i)}, \left\{ {X}^B_k +h \sum_{(j)=1}^{s} a_{(i)(j)}{K}^B_{(j)}\right\},  \left\{ {K}^B_{(i)} \right\}  \right) = 0. 
\]
Thanks to SDIRK Butcher tableau (\ref{tab:SDIRKtableau}), at a given Runge-Kutta step $(i)$, the quantity
\[
M^A_{(i)}  =X^A_k +h \sum_{(j)=1}^{(i)-1} a_{(i)(j)}{K}^A_{(j)}
\]
is known and therefore the ellipitc equation determining  $K^A_{(i)}$ is
\beq
 F^A\left(\tau_k + hc_{(i)}, \left\{ {M}^B_{(i)} + h\gamma{K}^B_{(i)}\right\},  \left\{ {K}^B_{(i)} \right\}  \right) = 0.
\eeq 
\subsection{The Bi-Conjugate Gradient Stabilized Method}\label{app:BiCGStab}

A systematic description of the BiCGStab-method is available in many books (see e.g, \cite{saad_96,Meister99}). We reproduce here the steps of the algorithm in order to adapt it to the notation introduced in section \ref{sec:SpecCode}. First we recall that for a given iteration $n$ of the Newton-Raphson scheme, we have to solve the linear system (\ref{eq:NR_linSys})
\[
\displaystyle
\underbrace{\left[ \frac{\partial E^A}{\partial X^B} + \frac{\partial E^A}{\partial X^B_{,\mu}}\frac{\partial}{\partial x^\mu} +  \frac{\partial E^A}{\partial X^B_{,\mu \nu}}\frac{\partial^2}{\partial x^\mu\partial x^\nu}\right]}_{\displaystyle J_B^{A}}\delta X^{B}+ {E}^{A}=0.
\]
Eq.~(\ref{eq:NR_linSys2}) expresses the linear system with the shorter notation
\[
L^A\left(\tau, \left\{\delta X^{B}\right\}, \left\{\partial_\tau {\delta X}^{B}\right\}, \left\{\partial^2_{\tau\tau} {\delta X}^{B}\right\}; X^{B}, E^{B}\right) = 0.
\]
Note that eqs. (\ref{eq:NR_linSys}) and (\ref{eq:NR_linSys2}) have an index $n$ indicating the specific Newton-Raphson step, which is omitted here in order to avoid confusion with the iterations within the BiCGStab-method.

For the detailed description of the BiCGStab-method, it is convenient to introduce further
\beq
{\cal J}^A\left(\tau, \left\{\delta X^{B}\right\}, \left\{\partial_\tau {\delta X}^{B}\right\}, \left\{\partial^2_{\tau\tau} {\delta X}^{B}\right\}; X^{B}\right) = J_B^{A}\delta X^{B}.
\eeq 

For a given initial guess ${\delta X}^{A}{}_{(0)}$ the algorithm reads as follows (we use Einstein's convention, see footnote \ref{footnote:EinsteinConv}): 
\ben
\item $\displaystyle r^A{}_{(0)} = L^A\left(\tau, \left\{\delta X^{B}{}_{(0)}\right\}, \left\{\partial_\tau {\delta X}^{B}{}_{(0)}\right\}, \left\{\partial^2_{\tau\tau} {\delta X}^{B}{}_{(0)}\right\}; X^{B}, E^{B}\right)$

\item  $\displaystyle \hat{r}^A{}_{(0)}=r^A{}_{(0)}$

\item  $\displaystyle \rho_{{0}} = \alpha = \omega_{(0)} = 1$, $v^A{}_{(0)} = p^A{}_{(0)} = 0$

\item for $i=1,2,3...$ 
	\ben
	\item $\displaystyle \rho_{(i)} = \hat{r}^A{}_{(0)}r_A{}_{(i-1)}$
	
	\item $\displaystyle \beta = \dfrac{\rho_{(i)}}{\rho_{(i-1)}}\dfrac{\alpha}{\omega_{(i-1)}}$
	
	\item $\displaystyle p^A{}_{(i)} = r^A{}_{(i-1)} + \beta\left( p^A_{(i)} -\omega_{(i-1)}v^A{}_{(i-1)}  \right)$
	
	\item solve $\displaystyle L^A\left(\tau, \left\{y^B\right\}, \left\{\partial_\tau y^B\right\}, \left\{\partial^2_{\tau\tau} y^B\right\}; X^{B}, p^{B}{}_{(i)}\right) = 0$ for $y^B$ \label{precond1}
	
	\item $\displaystyle v^A{}_{(i)} = {\cal J}^A\left(\tau, \left\{y^{B}\right\}, \left\{\partial_\tau y^{B}\right\}, \left\{\partial^2_{\tau\tau} y^{B}\right\}; X^{B}\right)$
	
	\item $\displaystyle \alpha = \dfrac{\rho_i}{\hat{r}^A{}_{(0)}v_A}$
	
	\item $\displaystyle s^A = r^A{}_{(i-1)} - \alpha v^A{}_{(i)}$
	
	\item solve $\displaystyle L^A\left(\tau, \left\{z^B\right\}, \left\{\partial_\tau z^B\right\}, \left\{\partial^2_{\tau\tau} z^B\right\}; X^{B}, s^{B}\right) = 0$ for $z^B$ \label{precond2}
	
	\item $\displaystyle t^A = {\cal J}^A\left(\tau, \left\{z^{B}\right\}, \left\{\partial_\tau z^{B}\right\}, \left\{\partial^2_{\tau\tau} z^{B}\right\}; X^{B}\right)$
	
	\item $\displaystyle \omega{}_{(i)} =(t^As_A)/(t^At_A)$
	
	\item $\displaystyle \delta X^{B}{}_{(i)} = \delta X^{B}{}_{(i-1)} + \alpha y^A + \omega_{(i)}z^A$
	
	\item if $\displaystyle L^A\left(\tau, \left\{\delta X^{B}{}_{(i)}\right\}, \left\{\partial_\tau {\delta X}^{B}{}_{(i)}\right\}, \left\{\partial^2_{\tau\tau} {\delta X}^{B}{}_{(i)}\right\}; X^{B}, E^{B}\right) < \delta$ for a given tolerance $\delta$ then quit
	\item else $\displaystyle r^A{}_{(i)}=s^A - \omega_{(i)}t^A$.
	\een
\een

The pre-conditioner corresponds to the steps (\ref{precond1}) and (\ref{precond2}). As described at the end of section \ref{sec:SpecCode}, it consists of an approximate solution for a linear system of the type (\ref{eq:BiCGstab_linSys}), which is obtained with the SDIRK-method introduced in section \ref{sec:SDIRK}.

\bibliographystyle{model1-num-names.bst}

\bibliography{bibitems}
\end{document}